\documentclass[aps,pra,twocolumn,showpacs,superscriptaddress,groupedaddress]{revtex4-1}

\usepackage[T1]{fontenc}   
\usepackage[utf8]{inputenc}

\usepackage{graphicx,color}
\expandafter\let\csname equation*\endcsname\relax
\expandafter\let\csname endequation*\endcsname\relax
\usepackage{amsmath}
\usepackage{amssymb}
\usepackage{hyperref}
\usepackage{bbold}
\usepackage{verbatim}
\usepackage{booktabs}
\usepackage[normalem]{ulem}

\def\ri{\mathrm i}
\def\re{\mathrm e}
\def\rd{\mathrm d}
\def\hc{{\rm h.c.}}

\begin{document}

\title{Drude weight increase by orbital and repulsive interactions in  fermionic ladders}

\author{Andreas Haller$^{1}$,  Matteo Rizzi$^{2,3}$ and  Michele Filippone$^4$}
\address{$^1$Institute of Physics, Johannes Gutenberg University, D-55099 Mainz, Germany}
\address{$^2$Institute of Quantum Control (PGI-8), Forschungszentrum J\"ulich, D-52425 J\"ulich, Germany}
\address{$^3$Institute for Theoretical Physics, University of Cologne, D-50937 K\"oln, Germany}
\address{$^4$Department of Quantum Matter Physics, Ecole de Physique University of Geneva, Quai Ernest-Ansermet 24, CH-1211 Geneva 4, Switzerland}


\begin{abstract}
In strictly one-dimensional systems, repulsive interactions tend to reduce particle mobility on a lattice. Therefore, the Drude weight, controlling the divergence at zero-frequency of optical conductivities in perfect conductors, is lower than in non-interacting cases. We show that this is not the case when extending to quasi one-dimensional ladder systems. Relying on bosonization, perturbative and matrix product states (MPS) calculations, we show that nearest-neighbor interactions and magnetic fluxes provide a bias between back- and forward-scattering processes, leading to linear corrections to the Drude weight in the interaction strength. As a consequence, Drude weights counter-intuitively increase (decrease) with repulsive (attractive) interactions. Our findings are relevant for the efficient tuning of Drude weights in the framework of ultracold atoms trapped in optical lattices and equally affect topological edge states in condensed matter systems.
\end{abstract}

\maketitle

\section{Introduction}\label{section:intro}
The seminal work by Walter Kohn~\cite{kohn_theory_1964} established the Drude weight as a crucial quantity to describe  the conduction properties of strongly correlated quantum systems. At zero temperature ($T\rightarrow0$), it weights the zero-frequency  divergence of  the conductivity~\cite{shastry_twisted_1990,fye_drude_1991,scalapino_superfluid_1992,giamarchi_persistent_1995}
\begin{equation}\label{eq:sigma}
\mbox{Re}\big[\sigma(\omega)\big]=\mathcal D\delta(\omega)+\sigma_{\rm reg}(\omega)\,,
\end{equation}
signaling  a  perfect conductor, which, as conventional superconductors,  supports non-dissipative/ballistic transport. In quasi one-dimensional conducting rings of size $L$, the Drude weight also determines the dissipationless persistent current $J=\mathcal D\Phi/L$ ~\cite{buttiker_josephson_1983,levy_magnetization_1990,bleszynski-jayich_persistent_2009,kulik_persistent_2010,viefers_quantum_2004} generated  in response to a (infinitesimal) magnetic flux $\Phi$ threading the ring. Persistent currents are an equilibrium property of quantum coherent conductors and are a phase-coherent manifestation of the Aharonov-Bohm  (AB) phase $\Phi$ acquired by particles upon looping around the ring, equivalent to a twist in the periodic boundary condition. As a consequence, Drude weights coincide with the susceptibility of the ground-state energy $E$ to such a twist:
\begin{equation}\label{eq:drudetwist}
\mathcal D=L\pi\left.\frac{\partial^2 E}{\partial \Phi^2}\right|_{\Phi\rightarrow0}\,.
\end{equation}
In conducting systems, the Drude weight remains finite in the thermodynamic limit, while its exponential suppression signals insulating behavior.
Beyond its  usefulness for  analytical and numerical calculations, Eqs.~\eqref{eq:sigma} and~\eqref{eq:drudetwist} establish a remarkable connection between the transport properties and the sensitivity to modified boundaries of quantum-coherent systems, underpinning, for instance, the scaling theory of Anderson localization~\cite{thouless_electrons_1974,edwards_numerical_1972,abrahams_scaling_1979,akkermans_conductance_1992,bouzerar_persistent_1994} and many-body generalizations thereof~\cite{von_oppen_exact_1994,von_oppen_exact_1995,filippone_drude_2016}.

Recent interest in  Drude weights is motivated by its rich behavior displayed in the presence of interactions and the general importance for experiments addressing novel quantum  transport phenomena: Surprisingly, the divergent contribution in Eq.~\eqref{eq:sigma} does not always disappear at finite temperature in fine-tuned integrable models~\cite{zotos_finite_1999,rosch_conductivity_2000,prosen_open_2011} and Drude weights contribute to  the Hall response of quasi one-dimensional (1D) systems~\cite{zotos_reactive_2000,greschner_universal_2019}.

 Importantly, synthetic quantum matter systems, such as ultracold atoms confined in ring-shaped optical traps~\cite{sauer_storage_2001,gupta_bose-einstein_2005,ryu_observation_2007,lesanovsky_time-averaged_2007,eckel_interferometric_2014,lacki_quantum_2016,amico_2005,cominotti_2014,gallemi_2018}, provide an experimental platform to study orbital responses to an applied flux in which temperature, particle statistics, and even interactions can be engineered almost at will. Moreover, the currents driven by either displacing the confining potential~\cite{mancini_observation_2015} or tilting the system~\cite{genkina_imaging_2019} reproduce those generated persistently by a flux $\Phi$ in a ring geometry in an adiabatic approximation~\cite{greschner_universal_2019}, therefore accessing Drude weights with open boundary conditions.

It is thus important, both on the experimental and fundamental level, to understand and develop physical intuition concerning the effects of strong correlations on the Drude weight.  It is commonly believed that repulsive interactions reduce particle mobility in a many-body system, leading to a generic reduction of the Drude weight \cite{dias_persistent_current_2006,meden_persistent_currents_2003,bouzerar_persistent_1994,berkovits_influence_1993}.
Nevertheless, it was recently observed that this fact is remarkably violated in Creutz ladders~\cite{bischoff_tuning_2017}:
this was attributed to the presence of an isolated Dirac cone and put into relation to the anomalous magnetic orbital response of 2D graphene~\cite{principi_drude_graphene_2010}.

In this work, we show that the increase of Drude weight by local repulsive interactions at zero temperature is actually a more general
feature of quasi-1D interacting systems thread by a {\it transverse} magnetic flux $\chi$, see Fig.~\ref{fig:hamiltonian_picture}.
We show this phenomenon by first relying on perturbative calculations, which are nicely reproduced by matrix product states (MPS) simulations.
We understand this phenomenon by deriving the effective low-energy Luttinger Liquid theory of interacting quantum ladders,
showing that nearest-neighbor interactions and magnetic fluxes provide a bias between back- and forward-scattering processes.
Finally, we connect this phenomenon with the Quantum Hall effect~\cite{bernevig_topological_2013},
in which the presence of magnetic fluxes leads to the suppression of interaction-induced backscattering.
As a consequence, the prominence of forward-scattering on polarized edge states leads to increased mobility,
signalled by an increased Drude weight.

Remarkably, such Drude weight corrections are {\it linear} in the interaction strength, thus allowing for efficient tunability of this quantity by switching to attractive interactions, in which case the Drude weight is suppressed.
Our findings are important as they shed new light on the transport properties of strongly correlated systems with non-trivial topological properties, which may be accessed both in synthetic and solid state quantum matter systems.

The work is structured as follows:
In Section~\ref{section:drude_increase}, we review why repulsive density-density interactions do not yield {\it strong}
(i.e. linear in the interaction strength) renormalization of the Drude weight in {\it strictly} one-dimensional systems.
Section~\ref{section:interacting_creutz} reviews the single-particle properties and band structure of the Creutz model. In Sec.~\ref{section:pert}, we calculate the  leading order corrections to the Drude weight in the interaction strength, showing quantitative agreement to MPS simulations. Section~\ref{section:bosonization} presents the effective Luttinger Liquid interpretation of the Drude weight increase, showing bias between back- and forward-scattering processes induced by orbital effects, making connection to Quantum Hall regimes in two dimensions.


\section{Weak Drude weight  renormalization in 1D}\label{section:drude_increase}

In strictly one dimension, interactions cannot possibly affect the Drude weight in Galilean invariant systems,
due to the perfect decoupling of the center-of-mass motion (affected then by the flux insertion)
from the internal degrees of freedom (affected instead by interactions).
Noticeably, this holds true also for multi-component systems, as long as the only coupling between different species is of density-density nature:
in that case, it will be the total current (and therefore the total Drude weight) to be untouched by interactions,
while off-diagonal drag coefficients may depend on the inter-species interaction strength.
The presence of a lattice, once away from commensurate effects which might resonate and open a gap,
is expected to affect this important result only beyond leading order.
These facts are readily understood in the bosonization formalism~\cite{giamarchi_quantum_2004},
which will later help us to clarify where the hack in the ladder case resides.

Consider a generic single-band tight-binding Hamiltonian,
$\mathcal H_{\rm kin}=\sum_k\varepsilon(k)n_k$, in which $\varepsilon(k)$ is the band dispersion and $n_k$ the density at momentum $k$.
In a low-energy approximation, the band dispersion $\varepsilon(k)$ can be linearized close to the Fermi points,
$\varepsilon(k)\sim\varepsilon(\pm k_{\rm F})\pm v_\mathrm{F}(k\mp k_{\rm F})$,
with $v_\mathrm{F}$ the Fermi velocity.
Such linearization allows to define two separate  right- and left-moving  fermions ($R_j/L_j$) out of the original fermions
on lattice site $j$, $c_j\sim e^{ik_{\rm F}j}R_j+e^{-ik_{\rm F}j}L_j$,
and their densities $n_{\alpha,i}=\alpha^\dagger_i\alpha_i^{\vphantom\dag}$ ($\alpha=R,L$).
In such ``Tomonaga-Luttinger-Dirac'' approximation,
the fermionic fields can in turn be expressed via a pair of canonically conjugate bosonic fields $\phi$ and $\theta$,
$[\phi(x),\partial_{x'}\theta(x')]=i\pi\delta(x-x')$, which describe density and current fluctuations of the effective low-energy system:
$n_i=n_{R,i}+n_{L,i}= -\partial_x\phi(x)/\pi$ and $J_i/v_\mathrm{F}=n_{R,i}-n_{L,i}= \partial_x\theta(x)/\pi$, respectively.
The Hamiltonian can be then exactly cast in the  bosonized form
	$\mathcal  H_{\rm kin} 
		=\int\frac{d x}{2\pi} v_\mathrm{F}\left[(\partial_x\theta(x))^2 +  (\partial_x\phi(x))^2\right]$%
~\cite{giamarchi_quantum_2004,delft_bosonization_1998}.
%
%
%
%
In the presence of (short-range) interactions, described by the Hamiltonian $\mathcal H_{\rm int}=\sum_{i,j}V_{i,j}n_in_j$,
the Luttinger Liquid  Hamiltonian is only slightly modified to
\begin{align}\label{eq:ll}
  \mathcal  H_{\rm LL} =\int\frac{d x}{2\pi} \left[uK\Big(\partial_x\theta(x)\Big)^2 +  \frac uK \Big(\partial_x\phi(x)\Big)^2\right]+\mathcal V[\phi]\,,
\end{align}
in which the Luttinger parameters $u$ and $K$ correspond to  the velocity of the collective plasma oscillation of the gas and give information about interactions, respectively: e.g., repulsive 1D fermions are usually described by $K \leq 1$.
The additional term $\mathcal V[\phi]$ collects all additional non-quadratic terms generated by density-density interactions, which -- importantly -- do not depend on the bosonic operator $\theta$, be the system on a lattice or not:
Indeed, the definition of the current operator obtained via the continuity equation is unaffected,
\begin{eqnarray}
\nonumber
[n,\mathcal H_{\rm int}] = 0 \  \Longrightarrow \
\partial_x J & = & -\partial_t n=-i[n,\mathcal H]=-i [n,\mathcal H_{\rm kin}] \\
& = & i[\partial_x\phi,\mathcal H_{\rm kin}]/\pi=v_\mathrm{F}\partial_x^2\theta/\pi .
\label{eq:nHcomm}
\end{eqnarray}%
As a crucial consequence, the product $uK$ -- i.e., the Drude weight -- remains equal to the non-interacting Fermi velocity $v_\mathrm{F}$, unaffected by interactions!
Standard minimal coupling, in which threading a flux $\Phi$ in a ring geometry is equivalent to shift
momenta as $k\rightarrow k-\Phi/L$ (i.e., here $\partial_x\theta(x)\rightarrow \partial_x\theta(x)-\Phi/L$),
combined with Eqs.~\eqref{eq:drudetwist} and~\eqref{eq:ll}, leads to
\begin{equation}\label{eq:DuK}
\mathcal D=uK=\mathcal D_0=v_\mathrm{F}\,.
\end{equation}%
In the absence of commensurability effects or other gap-formation mechanisms,
the corrections $\mathcal V[\phi]$ are usually irrelevant in the renormalization group sense~\cite{giamarchi_quantum_2004},
and therefore they do not affect the validity of the Hamiltonian~\eqref{eq:ll}
but at most lead to a renormalization of the Luttinger Liquid parameters $(u,K)\rightarrow (u^*,K^*)$.
The Drude weight gets renormalized $\mathcal D_0=v_\mathrm{F}\rightarrow \mathcal D^*=u^*K^*$ as well, but
-- crucially for the following discussion -- such renormalization is usually {\it weak} (i.e., at most of order $V^2$ in the perturbative expansion)
and the Drude weight is suppressed $\mathcal D^*<\mathcal D_0$%
~\cite{bouzerar_persistent_1994,meden_persistent_currents_2003,dias_persistent_current_2006}.

Alternatively, as we are going to rederive in the following, the validity of the identity~\eqref{eq:DuK} is also understood from the fact that interactions generate coupling between left- and right-movers -- so-called $g_2$ backscattering processes:   $g_2n_Rn_L$ -- and right/left movers themselves -- the so-called $g_4$ forward-scattering processes: $g_4(n_Rn_R+n_Ln_L)/2$. The Drude weight is affected by them as follows~\cite{giamarchi_quantum_2004}
\begin{equation}\label{eq:Dgg}
\mathcal D=\mathcal D_0 +\frac{g_4-g_2}{2\pi}\,,
\end{equation}
and, for the same reasons leading to Eq.~\eqref{eq:DuK}, in conventional lattice systems one always finds $g_4=g_2$ and thus no renormalization of the Drude weight occurs to leading order in the interactions.

The arguments leading to the identity~\eqref{eq:DuK}, namely the commutation rule~\eqref{eq:nHcomm}, do not generally apply in the presence of orbital effects in quasi-one dimensional systems and, in this paper, we show a very simple mechanism leading to a modification of the Drude weight in quantum ladders which is
{\it linear} in the interaction amplitudes and, remarkably, is {\it positive} in presence of typical intra-chain repulsive terms.


\section{The model}\label{section:interacting_creutz}

\begin{figure}[t]
    \centering
    \includegraphics[width=\columnwidth]{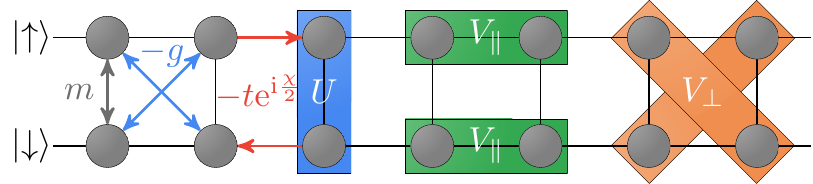}\llap{\parbox[b]{4.6cm}{(a)\\\rule{0ex}{.8cm}}}\\[.25cm]
    \includegraphics[width=\columnwidth]{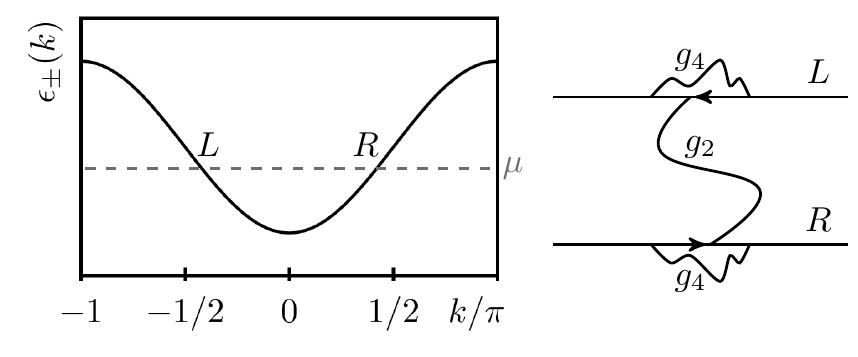}\llap{\parbox[b]{4.6cm}{(b)\\\rule{0ex}{3cm}}}\\[.25cm]
    \includegraphics[width=\columnwidth]{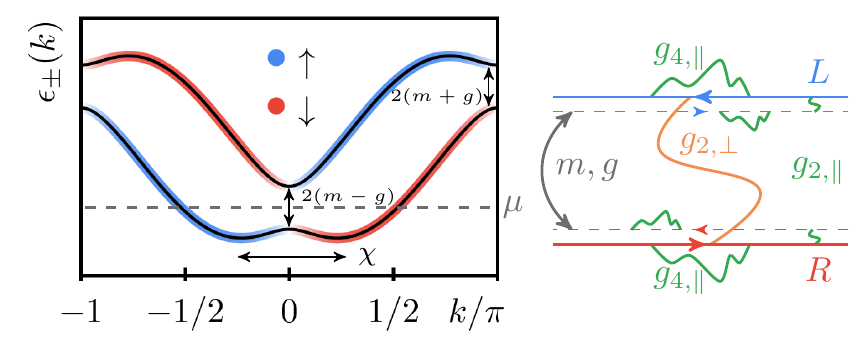}\llap{\parbox[b]{4.6cm}{(c)\\\rule{0ex}{3cm}}}\\[.25cm]
    \caption{(a) The interacting Creutz model, Eqs.~(\ref{eq:hamiltonian_noninteracting}-\ref{eq:hU}). We represent interactions by colored boxes and tight-binding hoppings as arrows connecting different lattice sites. Panel (b) depicts the usual one-dimensional case. A single pair of distinct left- and right-movers is identified close to the chemical potential $\mu$ and $V_\parallel$ induces $g_{2/4}$ interaction processes that exactly compensate each other on the lattice. Panel (c) illustrates the 2-leg ladder case with magnetic flux ($g=0$ for simplicity). A finite inter-chain hopping of amplitude $m$ gaps out only one among the right- or left-movers of each singular chain (dashed lines), leading to a suppression of backscattering $g_{2,\parallel}$ processes compared to forward scattering $g_{4,\parallel}$ processes caused by $V_\parallel$.
    As a consequence, the Drude weight increases. We also sketch how $V_\perp$ increases back-scattering, opposite to the action of $V_\parallel$.}
    \label{fig:hamiltonian_picture}
\end{figure}

The reference system and a sketch of the physical processes at work are illustrated in Fig.~\ref{fig:hamiltonian_picture}.
As an illustration, we consider a two-leg ladder of fermions (labeled as $\uparrow$ and $\downarrow$ species) where the plaquettes are threaded by a magnetic flux $\chi$.
The generalization to the  case with $N$ legs, relevant for topologically protected Quantum Hall regimes, is discussed in~\ref{app:multi}, with similar conclusions.
For the kinetic/non-interacting part of the  Hamiltonian $\mathcal H_{\rm kin}$, we consider the following:
\begin{align} \label{eq:hamiltonian_noninteracting}
    \mathcal H_{\rm kin} = \frac12\sum_{j=1}^Lc^\dag_{j\vphantom1}\left(\left(-t\re^{\ri\frac\chi2\sigma_z} - g\sigma_x\right)c_{j-1} + m \sigma_x c_{j\vphantom1}\right)+\hc\,,
\end{align}
in which we assume periodic boundary conditions ${c_0=c_L}$, define the fermionic annihilation operators ${c_j=(c_{j\uparrow},c_{j\downarrow})^T}$ and express the tight-binding Hamiltonian using the Pauli matrices $\sigma_x,\sigma_y$ and $\sigma_z$.
For $m\neq$0, $g=0$ and generic values of $\chi$, this represents the minimal instance of a quasi-two dimensional system pierced by magnetic flux, which has been extensively investigated under various aspects~\cite{ledermann_phases_2000,feiguin_pair_correlations_2009,petrescu_chiral_states_2015,cornfeld_chiral_currents_2015,greschner_symmetry_breaking_2016,calvanese_strianti_laughlin_2017,petrescu_laughlin_2018,haller_resonance_2018}, but interestingly not the one addressed here.
For $\chi=\pi$ and $g \neq 0$, the model is the Creutz ladder~\cite{creutz_end_1999,junemann_exploring_2017}, at whose fine-tuned point $m=g=t$ the anomalous behaviour of the Drude weight was originally pointed out and attributed to the presence of an isolated Dirac cone~\cite{bischoff_tuning_2017}.

On top of the tight-binding part, we dress the lattice with orbital-selective density-density interactions ${\mathcal  H_\text{\rm int} = \mathcal H_\parallel +\mathcal  H_\perp+\mathcal H_U}$.
Our main focus will be on intra-chain (parallel) nearest-neighbors interactions
\begin{align} \label{eq:hamiltonian_interactions}
    \mathcal H_\parallel &= V_\parallel \sum_j\left( n_{j,\uparrow}n_{j+1,\uparrow} + n_{j+1,\downarrow}n_{j,\downarrow} \right)\,,
\end{align}
but, motivated by recent experiments achieving orbital effects with synthetic dimensions~\cite{mancini_observation_2015,genkina_imaging_2019}, we will also consider nearest-neighbor (perperdicular) and $SU(2)$ symmetric  on-site repulsion between different legs
\begin{align}
    \mathcal H_\perp &= V_\perp \sum_j\left( n_{j,\uparrow}n_{j+1,\downarrow} + n_{j+1,\downarrow}n_{j,\uparrow} \right) \,, \\
\label{eq:hU}    \mathcal H_U &= U \sum_j n_{j,\uparrow}n_{j,\downarrow}\,.
\end{align}

\begin{figure*}[t]
    \centering
    \includegraphics{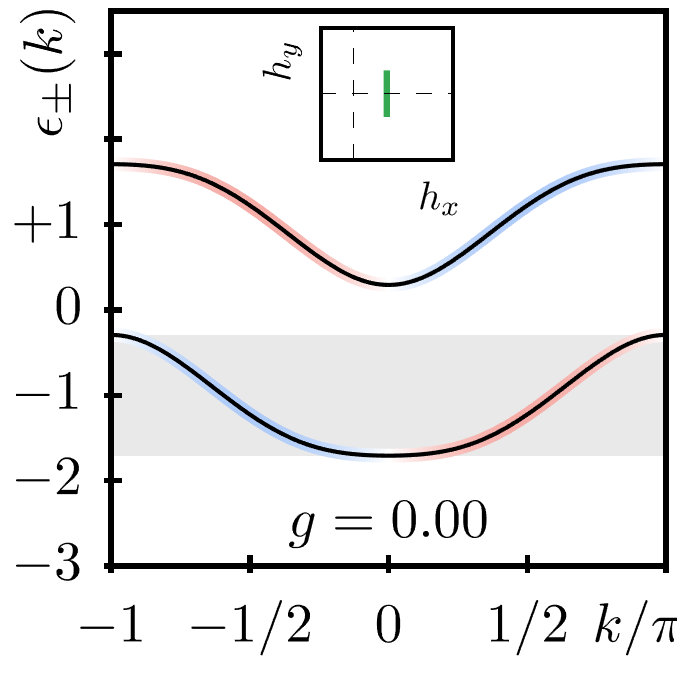}
    \includegraphics{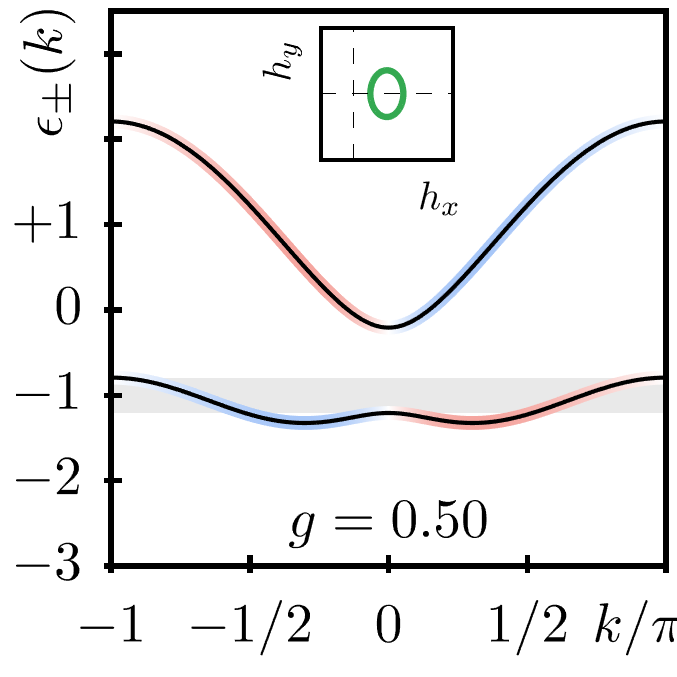}
    \includegraphics{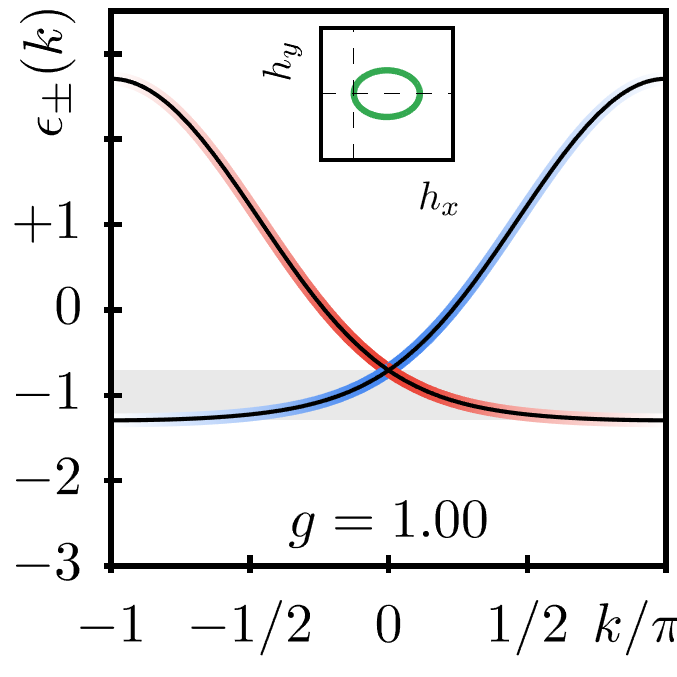}
    \includegraphics{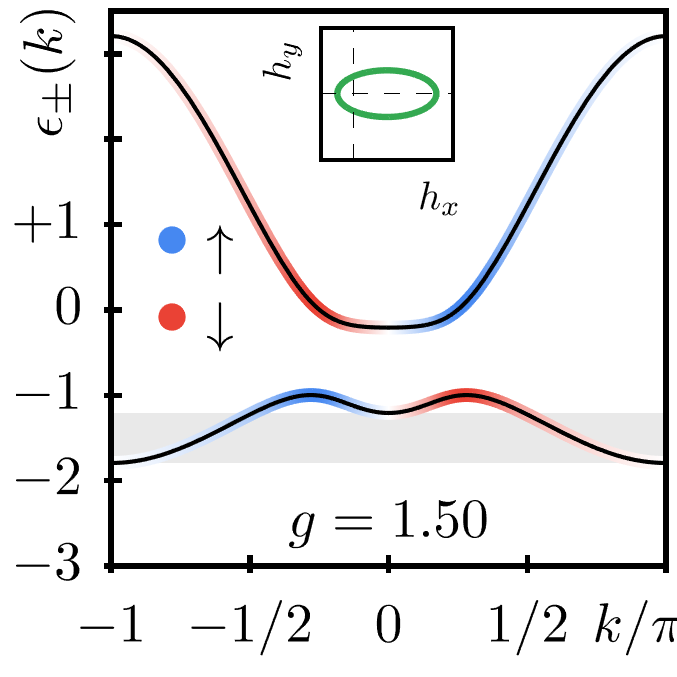}
    \caption{
        Spin components of the bands for different parameters $g,\chi$ as displayed at the top of the figures -- we fix $m=t=1$ and $\chi=\pi/2$ for simplicity.
        Black lines mark the band dispersion and the color represents the band polarization along the canonical axis ($\uparrow$ in blue, $\downarrow$ in red).
        White implies a perfect superposition of $\uparrow$ and $\downarrow$ species.
        The green inset circle represents the winding of $\vec h(k)$ and the dashed axes mark the origin.
        The gray region marks chemical potentials to fix a central charge $c=1$ with a single left and right mover at $\pm k_{\rm F}$ of the bottom band which serves as the starting point of the bosonization approach in Ch.~\ref{section:bosonization}.
        }
        \label{fig:dispersion}
\end{figure*}

Before considering the effect of interactions on the Drude weight, we discuss first the band-spectrum of the non-interacting model~\eqref{eq:hamiltonian_noninteracting}, which in Fourier $k$-space  reads
\begin{align}\label{eq:fourier}
    \mathcal H_{\rm kin} &= \sum_k c^\dag_k H_0(k) c^{\vphantom\dag}_k\,,& H_0(k)& = h_0(k)\mathbb1+\vec h(k)\cdot \vec\sigma\,,
\end{align}
where $\vec h = (h_x, h_y, h_z)$ and $\vec\sigma=(\sigma_x,\sigma_y,\sigma_z)$.
In the gauge chosen above,
\begin{align}
    h_0&=-t\cos (k)\cos\left(\frac\chi2\right)\,, & h_x&=m-g\cos (k)\,, \\  h_y&=0 \,, &  h_z &=t\sin (k)\sin\left(\frac\chi2\right)\,,
\end{align}
which is readily put in the diagonal form ${\mathcal H_{\rm kin} = \sum_{k,\nu=\pm} \varepsilon_\nu(k) d_{\nu,k}^\dag d^{\vphantom\dag}_{\nu,k}}$ by the transformation $c_k=U(k)d_k$ with, for $h_y=0$,
\begin{align}
    U(k) &= \frac1{\sqrt2}\left(\mathbb1\cdot\sqrt{1+\frac{h_z}h}-\ri\,\mbox{sgn}[h_x]\sigma_y\cdot\sqrt{1-\frac{h_z}h}\right)\,,
\end{align}
leading to
\begin{align}
    \begin{pmatrix}
        c_{\uparrow,k} \\
        c_{\downarrow,k}
    \end{pmatrix}  &=
    \begin{pmatrix}
        u(k) & -v(k) \\
        v(k) & u(k)
    \end{pmatrix}
    \begin{pmatrix}
        d_{+,k} \\
        d_{-,k}
    \end{pmatrix}\,,\nonumber
\end{align}
\begin{align}\label{eq:bogoliubov}
    u &= \frac1{\sqrt2}\sqrt{1+{\tilde h}_z}\,, & v &= \frac{\mbox{sgn}[h_x]}{\sqrt2}\sqrt{1-{\tilde h}_z} \,,
\end{align}
with the dispersion ${\varepsilon_\pm(k) = h_0(k) \pm h(k)}$, norm of the Bloch vector ${h(k) =|\vec h(k)|=(h_x^2+h_y^2+h_z^2)^{-1/2}}$ and $\tilde h_i = h_i / h$.
This can be readily checked by verifying $U^\dagger(k) H_0(k) U(k) = h_0(k)\mathbb1 + h(k)\sigma_z$.
In Fig.~\ref{fig:dispersion}, different spectra are given for different set of parameters of the Creutz model, among which the band dispersion sketched in Fig.~\ref{fig:hamiltonian_picture} is reproduced.
In the following chapters, we will exploit heavily the two basic ingredients to obtain the strong renormalization of the Drude weight: (i) a transverse flux $\chi$ which polarizes the dispersion bands along a chosen axis in an {\it asymmetric} fashion and (ii) a gapping mechanism such that only one pair of the chiral modes remains intact.
As a consequence, the densities defined in the chosen axis (here, $\sigma_z$) are spread asymmetrically in $k$-space (as depicted in Figs.~\ref{fig:hamiltonian_picture}-\ref{fig:dispersion}) and same-spin density-density interactions favor forward-scattering, whereas different-spin density-density terms favor backscattering processes.

\section{Perturbation theory}\label{section:pert}
\begin{figure}[t]
    \includegraphics[width=.4925\columnwidth]{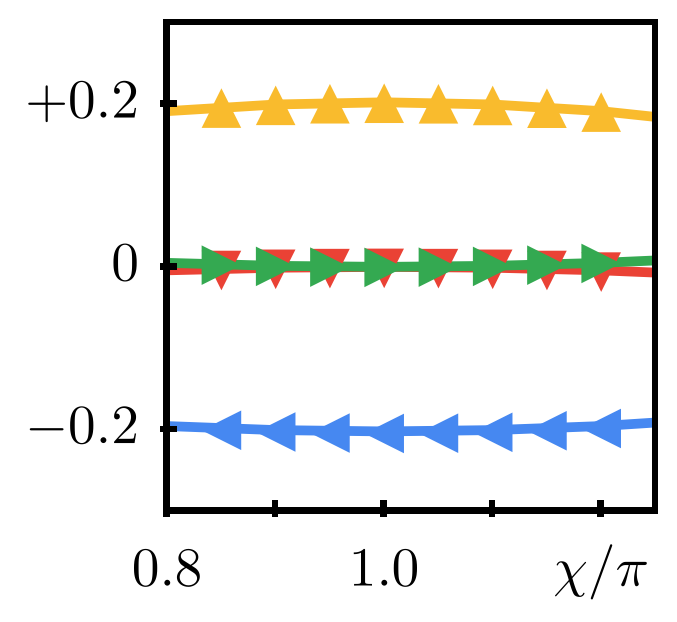}\llap{\parbox[b]{0.9cm}{(a)\\\rule{0ex}{2.6cm}}}
    \includegraphics[width=.4925\columnwidth]{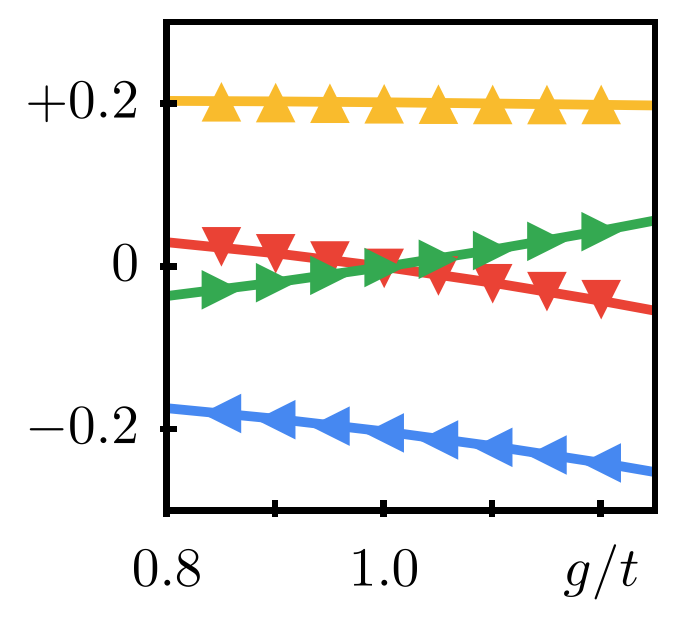}\llap{\parbox[b]{5.75cm}{(b)\\\rule{0ex}{2.6cm}}}
    \includegraphics[width=\columnwidth]{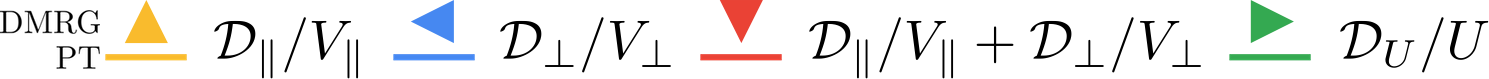}
    \caption{Renormalization of the Drude weight by interaction (different colors and symbols). Continuous lines represent the thermodynamic limit of the perturbative calculation given in Eqs.~\eqref{eq:pert_U},~\eqref{eq:pert_parallel},~\eqref{eq:pert_perp} and symbols represent results obtained by MPS simulations. Numerical values are converged up to the third decimal digit. The system size is $N/L=20/32$. Each set of data is obtained by the interplay of the kinetic term with individual interactions $\mathcal H = \mathcal H_{\rm kin} + \mathcal H_i$ with weak interaction amplitudes $V_\parallel=0.1t$, $V_\perp=0.1t$, $V_\parallel=V_\perp=0.1t$ and $U=0.125t$. The two figures demonstrate the robustness of the strong Drude weight renormalization, namely its increase for $V_\parallel>0$ and decrease for $V_\perp>0$ around the fine-tuned point of a single Dirac cone $m/t=g/t=\chi/\pi=1$. Hereby, we devote our focus to the influence of the individual interactions by subtracting the corresponding non-interacting value $\mathcal D_0$ from the total susceptibilxity, i.e. $\mathcal D_i = \mathcal D - \mathcal D_0$. In (a) we display the interaction-induced Drude weight shift $\mathcal D_i$ versus transverse flux $\chi$, in which the curvature of the two SU(2) symmetric interactions denoted by red (down-triangle) and green (right triangle) is not visible in the chosen scaling. To quantify, the plotted valus of $D_U/U$ are all inside the interval $[0,5]\cdot 10^{-3}$ and $\mathcal D_{\perp}/V_\perp + \mathcal D_{\parallel}/V_\parallel$ resides in $[-4.2,0]\cdot10^{-3}$. Panel (b) shows $\mathcal D_i$ as a function of the nearest neighbor inter-species hopping amplitude $g/t$.}\label{fig:pt_mps}
\end{figure}

As a supporting point for the Luttinger Liquid analysis we develop in Section~\ref{section:bosonization}, we first derive  the corrections to the Drude weight relying on standard  perturbation theory  to leading order in the interaction strengths $V_\parallel$, $V_\perp$, and $U$. Equation~\eqref{eq:drudetwist} requires to derive first the corresponding corrections to the ground-state energy. We focus on the situation of interest, in which only the lowest band $\varepsilon_-$ is occupied ($n_{+,k}=0\,\forall k$).
To leading order, the interaction-induced corrections to the Drude weight are obtained by averaging the interaction terms onto the ground state. A magnetic flux threading the ring is equivalent to twisting the boundary by a phase $\Phi$ and a matter of substituting $k\rightarrow k-\Phi/L$, upon which we expand to second order in $\Phi$ and then approach the thermodynamic limit $\frac 1L\sum_k \rightarrow \frac1{2\pi}\int\rd k$.
Exploiting the fact that $h_{x/z}$ is an even/odd function and the integral boundaries are all symmetric around $k=0$ ({\it after} the Taylor expansion in the flux $\Phi$), one finds
\begin{align}\label{eq:pert_parallel}
    \langle \mathcal H_\parallel \rangle &= \frac{V_\parallel L}{2}\left(n^2 - \left(\frac{H_{z,0}^{1,0}}{2\pi}\right)^2\right)+\frac{\Phi^2}{2\pi L}\mathcal D_\parallel\,,\nonumber\\
    \mathcal D_\parallel &= \frac{V_\parallel}{4\pi}\left({H^{0,0}_{z,1}}^2 - {H^{0,1}_{z,1}}^2 - H^{1,0}_{z,2}H^{1,0}_{z,0}\right)
\end{align}
with $n=N/L$ being the total density and we defined integral functions of the Bloch vector components $\tilde h_i$ which depend upon the Fermi sea ($FS$)
\begin{align}
    H_{i,n}^{\alpha,\beta}=\int_{FS}\left(\sin^\alpha(k)\cos^\beta(k)\frac{\partial^n}{\partial k^n}\tilde h_i(k)\right)\rd k\,.
\end{align}
The above and also all following expansions in the flux are actually correct up to $\mathcal O(\Phi^4)$ since all odd orders are proportional to symmetric integrals of odd functions and thus vanish.
The inter-species interaction returns
\begin{align}\label{eq:pert_perp}
    \langle \mathcal H_\perp \rangle &= \frac{V_\perp L}{2}\left(n^2 - \left(\frac{H_{x,0}^{0,1}}{2\pi}\right)^2\right)+\frac{\Phi^2}{2\pi L}\mathcal D_\perp\,,\nonumber\\
    \mathcal D_\perp &= -\frac{V_\perp}{4\pi}\left({H^{0,0}_{z,1}}^2 + {H^{1,0}_{x,1}}^2 + H^{0,1}_{x,2}H^{0,1}_{x,0}\right)
\end{align}
and the on-site interaction results in
\begin{align}\label{eq:pert_U}
    \langle\mathcal H_U \rangle &= \frac {UL}{4}\left(n^2-\left(\frac{{H_{x,0}^{0,0}}}{2\pi}\right)^2\right)+\frac{\Phi^2}{2\pi L}\mathcal D_U\,,\nonumber\\
    \mathcal D_U &= -\frac{U}{8\pi}\left({{H}^{0,0}_{z,1}}^2+H^{0,0}_{x,2}H^{0,0}_{x,0}\right)\,.
\end{align}
The Drude renormalization by interaction, according to leading order perturbation theory, depends on the precise form of the microscopic model and does not follow any universal law.
Strikingly, if we restrict to a {\em single} pair of Fermi points, the Drude weight {\em increases} for $V_\parallel>0$ and {\em decreases} for $V_\perp>0$ throughout the entire phase space $m,g,t$ which can be readily checked by evaluating Eqs.~\eqref{eq:pert_U},\eqref{eq:pert_parallel} and \eqref{eq:pert_perp} (see App.~\ref{app:perturbation_theory_details}, Fig.~\ref{fig:pert_results}).
As a trivial consequence, but in contrast to the common intuition, attractive interactions ($V_\parallel < 0$) decrease the orbital response function $\mathcal D$.
In case of absent transverse magnetic flux $\chi=0$, there is no renormalization of the Drude weight.
The  reason for such absence of strong (linear in the interactions) renormalization is the absence of any symmetry breaking mechanism between $g_4$ and $g_2$ processes induced by interactions, which are triggered by a finite $\chi$ as exemplified in Fig.~\ref{fig:hamiltonian_picture} based on the Luttinger Liquid analysis we carry out in Section~\ref{section:bosonization}.
Moreover, we also notice that the remarkable strong absence of renormalization  $\mathcal D_U=0$ and $D_\parallel/V_\parallel+D_\perp/V_\perp=0$ in leading order perturbation theory initially derived in~\cite{bischoff_tuning_2017} does not hold in general, but applies only for very special points in the phase space such as the single Dirac cone setting at $m/t=g/t=\chi/\pi=1$ (more generally, the particle-hole symmetric Dirac cone setting at $m/t=g/t= \sin(\chi/2)$).
Finally, we emphasize the perfect agreement between perturbative results and MPS simulations for a weakly interacting system of size $N/L=20/32$ as presented in Fig.~\ref{fig:pt_mps}.

\begin{figure}[t]
    \centering
    \includegraphics[width=.49\columnwidth]{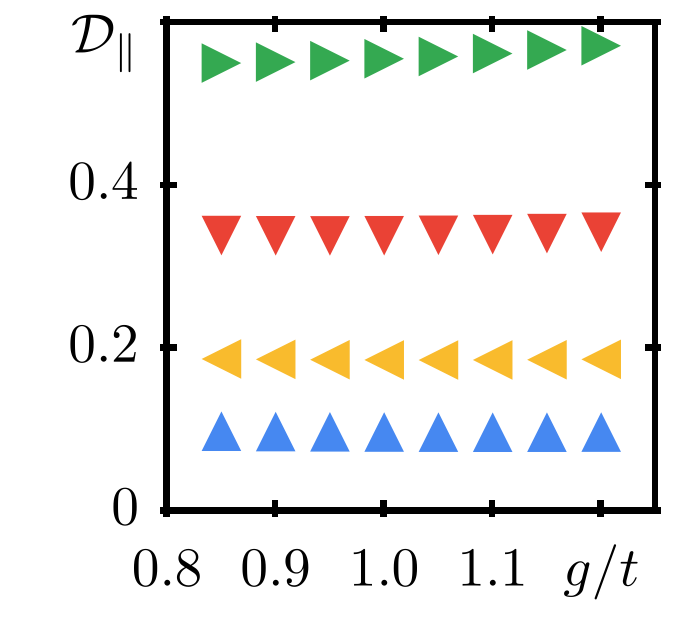}\llap{\parbox[b]{5.5cm}{(a)\\\rule{0ex}{1.8cm}}}
    \includegraphics[width=.49\columnwidth]{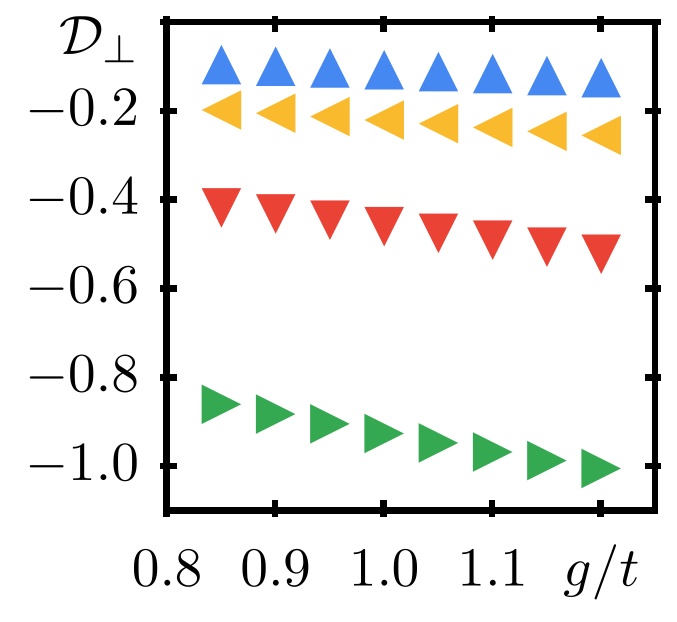}\llap{\parbox[b]{1cm}{(b)\\\rule{0ex}{1.8cm}}}
    \includegraphics[width=.49\columnwidth]{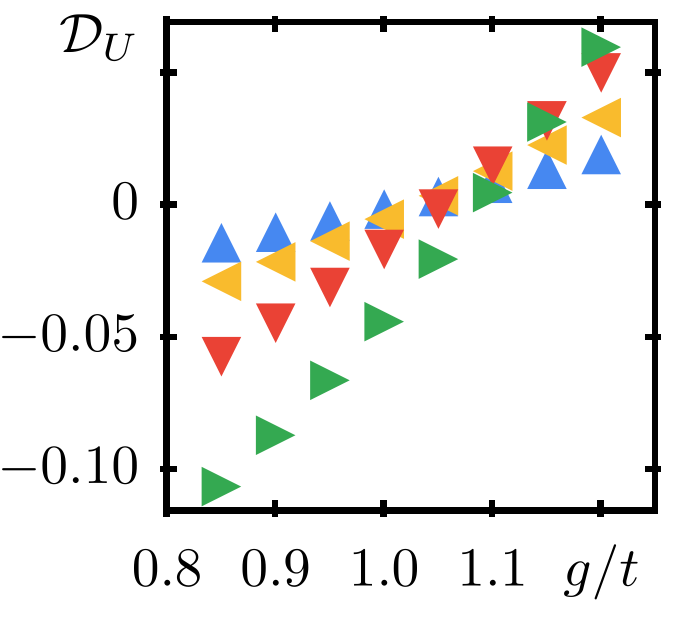}\llap{\parbox[b]{1cm}{(c)\\\rule{0ex}{0.85cm}}}
    \includegraphics[width=.49\columnwidth]{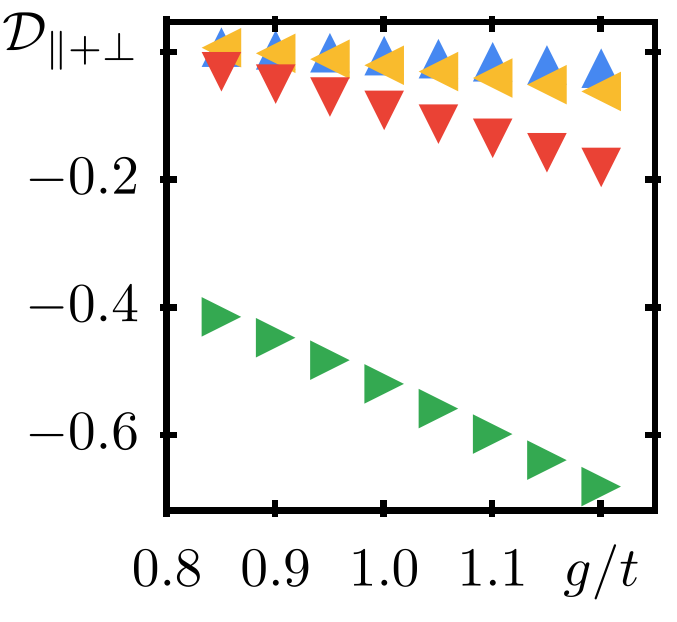}\llap{\parbox[b]{5.5cm}{(d)\\\rule{0ex}{0.85cm}}}
    \includegraphics[width=\columnwidth]{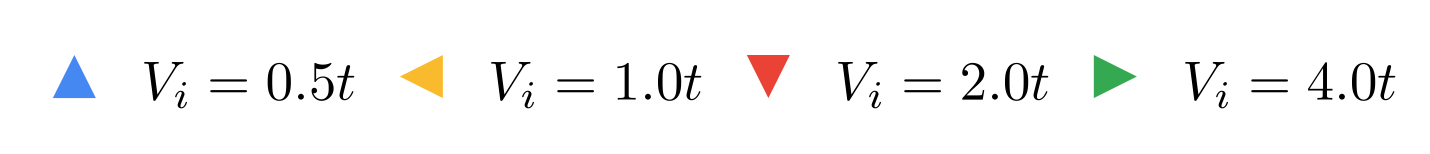}
    \caption{Change of the Drude weight obtained by MPS simulations for different interaction amplitudes signalled by different colors and markers. The displayed values are converged up to the third decimal digit. The density of the system is $n=\frac NL = \frac{20}{32}$, $m/t=\chi/\pi = 1$ and $g$ is relaxed to values around the topological transition at $g/t=1$. (a) Drude increase induced by $\mathcal H_\parallel$ and (b) Drude decrease by $\mathcal H_\perp$. (c) Drude renormalization by $\mathcal H_U$ and (d) by $\mathcal H_\parallel+\mathcal H_\perp$ for the case $V_\parallel=V_\perp$. The effect is observed for any simulated amplitude of the interactions and irrespective of the underlying topology. Notice that the renormalization of the Drude weight due to $\mathcal H_\perp$ is stronger and very asymmetrical compared to $\mathcal H_\parallel$.}
    \label{fig:drude_change_mps}
\end{figure}

We also stress the fact that such corrections to the Drude weight are actually {\it strong} and can be comparable to the non-interacting value itself, as clearly shown by the MPS simulations reported in Fig.~\ref{fig:drude_change_mps}, in which we considered values of the interactions comparable with the parameters of the non-interacting model~\eqref{eq:hamiltonian_noninteracting}.

The presented predictions from perturbation theory are qualitatively captured by the effective bosonized low-energy model which we derive in the next section.

\section{Bosonization and connection to Quantum Hall systems}
\label{section:bosonization}
The physical reason behind the linear increase and suppression of the Drude weight by the $V_\parallel$ and $V_\perp$ interaction, in the presence of a finite transverse flux $\chi$,  becomes apparent in the bosonization formalism. The bosonization of the interacting Creutz model requires a linearization of the spectrum close to the Fermi energy.

We consider the case of central charge $c=1$ (i.e., two Fermi points) with the chemical potential crossing the lower band $\varepsilon_-(k)$. Even though we keep the discussion general here, the reader can think of the situations depicted in Fig.~\ref{fig:dispersion}.
Proceeding in analogy to Refs.~\cite{narozhny_fractional_2005,carr_spinless_2006}, we  approximate
the two species of $\uparrow/\downarrow$ fermions as a superposition of a {\it single} pair of left ($L$) and right ($R$) movers.
Thus, the kinetic part of the Hamiltonian takes the form~\eqref{eq:ll} with $v_\mathrm{F}=\partial_k\varepsilon_-(k)|_{k=k_{\rm F}}$.
To bosonize the interactions, we switch to the continuum and apply the aforementioned transformation onto the spin-density operators
to be inserted in $\mathcal H_{\rm int}$. We consider situations out of quarter-filling, in which  Umklapp terms $\propto e^{4ixk_{\rm F}}R^\dagger R^\dagger L L$ stop oscillating and may cause relevant gap-leading perturbations~\cite{narozhny_fractional_2005}.
The total density operator thus becomes
\begin{eqnarray}
\nonumber n_i & \sim & n_R(x)+n_L(x) \\
& & - 2u(k_{\rm F})v(-k_{\rm F})[e^{-2ik_{\rm F}x}R^\dagger (x)L(x)+\mbox{h.c}].
\end{eqnarray}
This density representation differs from that of a truly 1D {spinless} Luttinger Liquid by the presence of the coherence factors $u$ and $v$.
At this stage it is possible to understand the reason why the relation~\eqref{eq:DuK}, valid for strictly 1D systems, does not hold in our context:
the projection via Eq.\eqref{eq:bogoliubov} on the low-energy sector captured by this bosonization approach is not a Bogoliubov transformation, as it does not preserve the fermionic anti-commutation relations of the operators $c_{\uparrow/\downarrow}$.
Moreover, it is responsible for the modification of the commutator with the interacting Hamiltonian which becomes non-zero, $[n(x),\mathcal H_{\rm int}]\neq 0$.
As we have already seen in Sec.~\ref{section:drude_increase}, such a condition is necessary not to modify the current operator and derive Eq.~\eqref{eq:DuK}.
Here, we explicitly lack this condition and expect a {\it strong} renormalization of the product $uK$ for the effective Hamiltonian.

The mapping to the standard Luttinger Hamiltonian~\eqref{eq:ll} occurs via the bosonization identities $R(x)\sim\re^{-\ri (\phi(x)-\theta(x))}$, $L(x)\sim\re^{\ri (\phi(x)+\theta(x))}$~\cite{giamarchi_quantum_2004}.
Reminding that ${n_R(x)=[\partial_x\theta(x)-\partial_x\phi(x)]/2\pi}$ and ${n_L(x)=-[\partial_x\theta(x)+\partial_x\phi(x)]/2\pi}$, we extract the bosonized Hamiltonian~\eqref{eq:ll} with renormalized Luttinger parameters
\begin{align} \label{eq:uk}
    \frac{uK}{v_\mathrm{F}} &=1+\frac1{2\pi}\sum_{i\in\{\parallel,\perp,U\}}V_i\left(g_{4,i}-g_{2,i}\right) \,,\\
    \frac u{v_\mathrm{F}K}&=1+\frac1{2\pi}\sum_{i\in\{\parallel,\perp,U\}}V_i\left(g_{4,i}+g_{2,i}\right) \,,
\end{align}
for which the detailed derivation of the $g$-factors is given in App.~\ref{app:bosonization_details}, with their explicit form in Table~\ref{tab:copuling_constants}.
The key result of this paper is resumed by the fact that
\begin{align}
    g_{4,\parallel}-g_{2,\parallel} &= +2(u^2-v^2)^2 = 2\tilde h_z(k_{\rm F})^2>0\,.
\end{align}
As a consequence, in the presence of repulsive intra-chain nearest-neighbor interactions, bosonization predicts the increase of the Drude weight by repulsive interactions.

Given the remarkable fact that such correction is {\it linear} in the interaction constant $V_{\parallel}$,  attractive interactions reduce the mobility of such systems ($\mathcal D<\mathcal D_0$).
Notice also  that this results holds irrespective of the topological nature of the bands in the Creutz model ($m\lessgtr g$).
The particular interest in the bosonization approach is that it allows to readily identify the breaking of symmetry between the relevant forward- and back-scattering  processes ($g_4$ and $g_2$ respectively) responsible for such mobility increase in the presence of intra-chain interactions.
Notice further that, also within bosonization, no renormalization of the Drude weight occurs in the absence of transverse magnetic flux $\chi=0$.
As sketched in Figure~\ref{fig:hamiltonian_picture}, the possibility to induce, via magnetic fluxes, orbital effects in such ladder system, allows to suppress interaction-induced backscattering between left-and right-movers as these modes are separated in space (spin-polarized).

Such a phenomenon can be put in connection to the exponential suppression of backscattering by topological bulk protection in Quantum Hall systems~\cite{bernevig_topological_2013}. As we discuss in detail by extending to the  multi-leg case in App.~\ref{app:multi}, following the spirit of the coupled-wire construction of topological insulators~\cite{kane_fractional_2002,meng_coupled-wire_2019}. As modes in the bulk are gapped, local interactions cannot efficiently couple any degree of freedom to the chiral modes which are localized  at the sample edge, thus backscattering is exponentially suppressed with the number of legs. Nevertheless a residual effect of interactions remains, which is forward scattering, that, as made explicit by the bosonization formula~\eqref{eq:Dgg}, increases the Drude weight. In Quantum Hall systems, which feature ballistic edges states at their border, the Drude weight is also expected to be renormalized by interactions~\cite{antinucci_universal_2018,wen_chiral_1990}. Nevertheless,  such corrections were never calculated explicitly in lattice models, especcially not the increase induced by repulsive interactions discussed in this work.
Our work predicts that, surprisingly, such renormalization is actually strong (linear) in the interaction strength and generally leads to an increase of the Drude weight for short-range repulsive interactions, in striking contrast to what expected in the one-dimensional limit.


Additionally, we notice that, in the bosonization formalism,  introducing inter-chain interactions $V_\perp$ has {\em exactly} the opposite effect on the Drude weight, namely its suppression. In the Quantum Hall picture discussed in the previous paragraph, the $\mathcal H_\perp$ interaction can be seen as a long-range interaction coupling left- and right-chiral edges. As a consequence,  backscattering is induced and thus  the Drude weight is reduced. This kind of long-range interaction is unlikely to affect condensed matter system. Nevertheless they (in particular $\mathcal H_U$) are actually the ones mainly at work in synthetic systems involving artificial gauge fields~\cite{mancini_observation_2015,genkina_imaging_2019}, whose are responsible of non-trivial effects~\cite{delre_2018} which deserve further investigation in the presence of orbital effects.


We conclude this section mentioning that the bosonization results are quantitatively different from the perturbation calculations reported in Sec.~\ref{section:pert}, on which we can fully rely given their perfect comparison with MPS calculations. Nevertheless the qualitative picture remains the same, apart from the perfect cancellation of the contributions resulting from $\mathcal H_{\parallel}+\mathcal H_{\perp}$. Such quantitative discrepancies are expected in bosonization given the strong approximation regarding the linearization of the dispersion and the presence of an underlying lattice in the microscopic model. In App.~\ref{app:bosonization_details},  these quantitative discrepancies are discussed in detail.

Moreover, we mention that the renormalization of the Drude weight of the $SU(2)$ symmetric interaction $\mathcal H_U$ remains intriguing.
In Figs.~\ref{fig:pt_mps}-\ref{fig:drude_change_mps}, $\mathcal D_U$ shows an interesting change of sign as a function of $g$.
Even though, also in this case, our perturbative calculations nicely reproduce the MPS simulations, we could not provide an intuitive explanation of this feature relying on bosonization.
The main difficulty here is also related to the lost of information about the fact that operators are originally at the same point after switching to the diagonal basis for Eq.~\eqref{eq:hamiltonian_noninteracting}.
We leave the investigation of this issue for future work.

\section{Discussion and Conclusion}

As mentioned in the Introduction,  there is a strong experimental and fundamental interest in the coherent transport properties of correlated systems confined to ring geometries~\cite{sauer_storage_2001,gupta_bose-einstein_2005,ryu_observation_2007,lesanovsky_time-averaged_2007,eckel_interferometric_2014,lacki_quantum_2016,amico_2005,cominotti_2014,gallemi_2018}. Knowledge about the behavior of the Drude weight in this context is crucial, as this quantity controls the magnitude of persistent currents in such systems.  
Our results shed new light on the transport properties of strongly correlated quantum systems, demonstrating how repulsive/attractive interactions counter-intuitively increase/decrease the mobility  of interacting fermions in the presence of orbital effects.

We derived the effective continuum Luttinger low-energy theory of the interacting Creutz model and focused on the situation with only two Fermi points, that is central charge $c=1$.
We have shown that the Drude weight changes linearly in the coupling parameters $V_\parallel, V_\perp$ of the two orbital-selective nearest-neighbor interactions.
Our study generalizes and clarifies the possibility of tuning the Drude weight observed in a previous study using MPS and second order perturbation theory~\cite{bischoff_tuning_2017}, predicting a linear dependence of the Drude weight with respect to $V_\parallel$, $V_\perp$ and $U$.
In this work, we limited the bosonization approach to the simplest situation with only two Fermi points, but a generalization to higher central charges is straightforward.
A direct comparison with leading order perturbation theory and MPS simulations reveals quantitative shortcomings of the standard bosonization procedure regarding the prediction of transport properties.
Future work in this direction should definitively address these issues.

Interesting perspectives concern the geometrical interpretation of the effects discussed in this work~\cite{hetenyi_2013} and the study of the interplay of quantum impurities and bulk interactions in such systems~\cite{sticlet_2013,cominotti_2014}.
Moreover, it is completely open for investigation how the interactions studied in this paper affect a setup in which energy and mass transport are induced by biased reservoirs~\cite{brantut_thermoelectric_2013,krinner_observation_2014,lebrat_2018,papoular_fast_2014,filippone_violation_2016,simpson_one-dimensional_2013}.
In this case, interactions  are not expected to affect the mass conductance~\cite{maslov_1995,*safi_1995,*ponomarenko_1995}, but only the thermal conductance, leading to the violation of the Wiedemann-Franz law~\cite{filippone_violation_2016}.
In this setting, transport is controlled by conductances, rather than Drude weights and transverse magnetic fields in the connecting region are expected to lead novel quantized effects~\cite{salerno_2019,filippone_2019}.
Another direction of interest would be to study the renormalization of the transverse flux $\chi$ susceptibility, in order to stabilize end enhance pretopological fractional excitations~\cite{calvanese_strianti_laughlin_2017,petrescu_laughlin_2018,haller_resonance_2018,strianti_pretopological_fqh_2019}.

\section{Acknowledgements}
The authors thank   C.-E. Bardyn, M. Burrello, P. van Dongen, T. Giamarchi, K. Le Hur, A. Minguzzi, S. Manmana, S. Paeckel, I. V. Protopopov, P. Schmoll and I. Schneider for fruitful and inspiring discussions.
A.H. is thankful for the financial support of the MAINZ Graduate School of Excellence and the Max Planck Graduate Center.
M.R. and A.H. acknowledge support from the Deutsche Forschungsgesellschaft (DFG) through the grant OSCAR 277810020 (RI 2345/2-1).
M.F. acknowledges support from the FNS/SNF Ambizione Grant PZ00P2\_174038.
The MPS simulations were run on the Mogon cluster of the Johannes Gutenberg-Universit\"at (made available by the CSM and AHRP), with a code based on a flexible Abelian Symmetric Tensor Networks Library, developed in collaboration with the group of S. Montangero at the University of Ulm (now moved to Padua).

\onecolumngrid
\appendix
\section{Perturbation theory -- supplementary}
\label{app:perturbation_theory_details}
\begin{figure}[ht]
    \centering
    \includegraphics[width=.24\columnwidth]{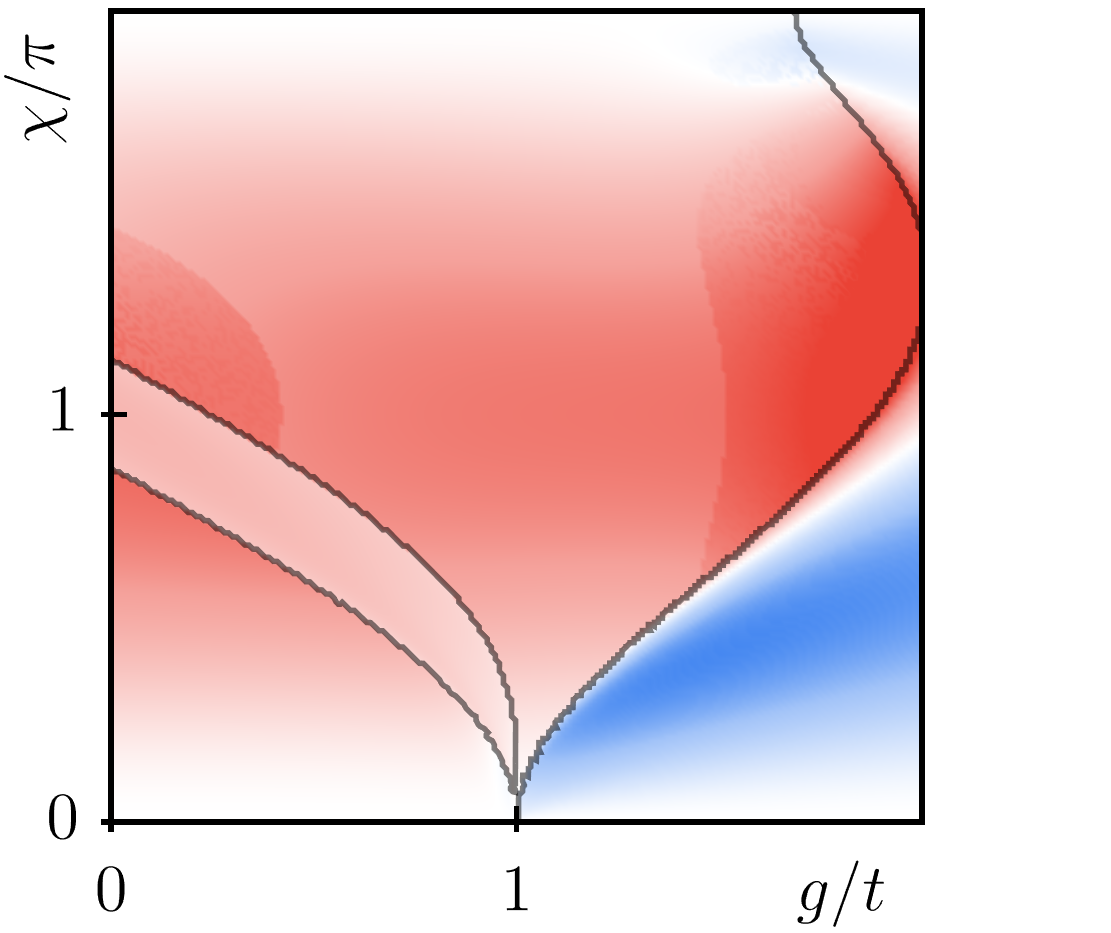}\llap{\parbox[b]{6.8cm}{(a)\\\rule{0ex}{2.95cm}}}\llap{\parbox[b]{4.8cm}{$c=1$\\\rule{0ex}{2cm}}}\llap{\parbox[b]{2.8cm}{$2$\\\rule{0ex}{1cm}}}\llap{\parbox[b]{6.cm}{$2$\\\rule{0ex}{1.35cm}}}\llap{\parbox[b]{6.5cm}{$1$\\\rule{0ex}{0.75cm}}}
    \includegraphics[width=.24\columnwidth]{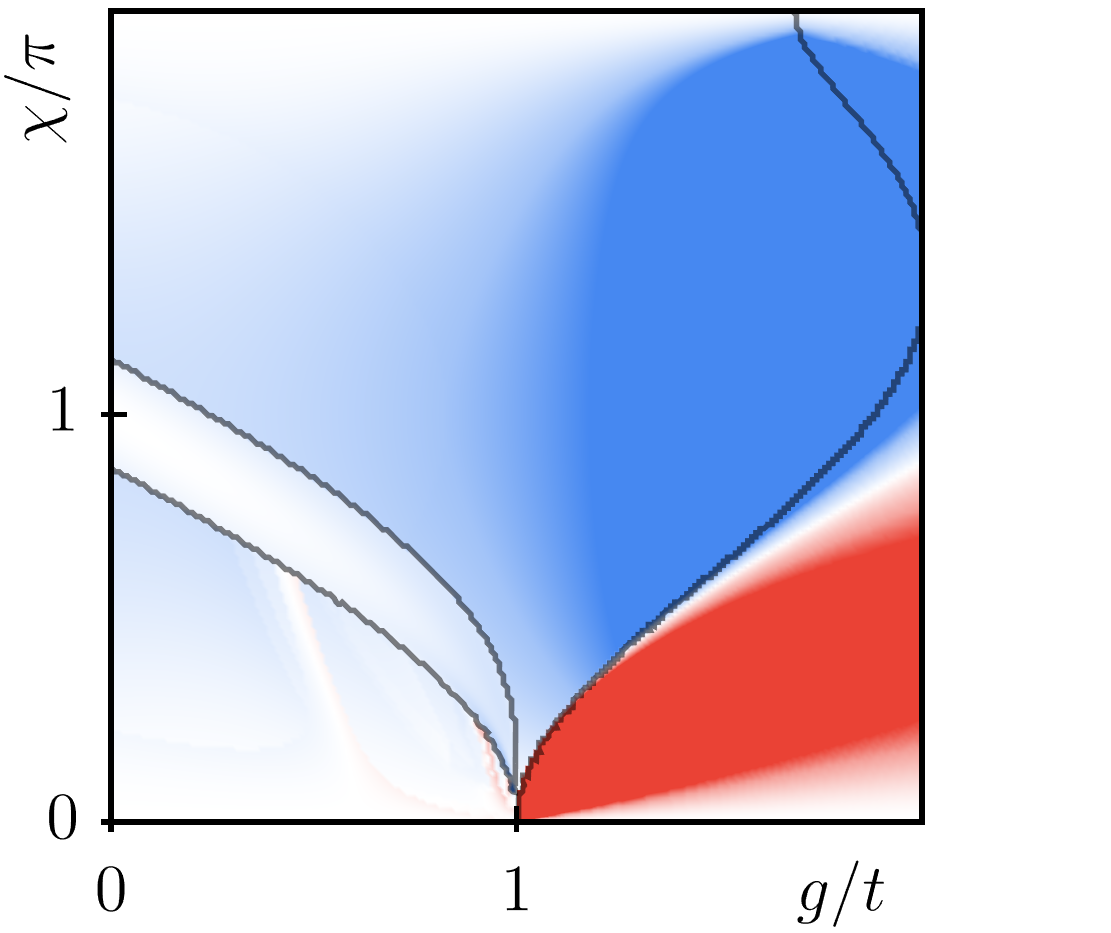}\llap{\parbox[b]{6.8cm}{(b)\\\rule{0ex}{2.95cm}}}\llap{\parbox[b]{4.8cm}{$c=1$\\\rule{0ex}{2cm}}}\llap{\parbox[b]{2.8cm}{$2$\\\rule{0ex}{1cm}}}\llap{\parbox[b]{6.cm}{$2$\\\rule{0ex}{1.35cm}}}\llap{\parbox[b]{6.5cm}{$1$\\\rule{0ex}{0.75cm}}}
    \includegraphics[width=.24\columnwidth]{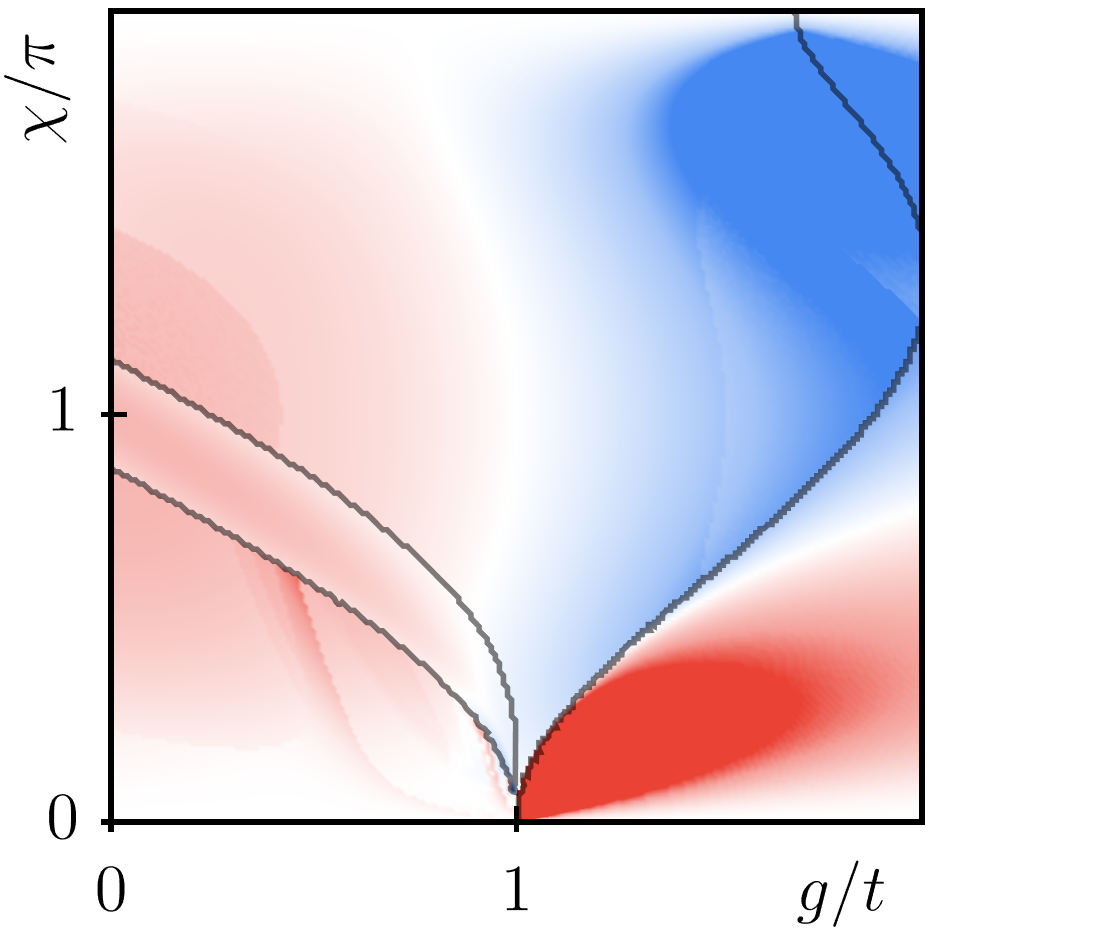}\llap{\parbox[b]{6.8cm}{(c)\\\rule{0ex}{2.95cm}}}\llap{\parbox[b]{4.8cm}{$c=1$\\\rule{0ex}{2cm}}}\llap{\parbox[b]{2.8cm}{$2$\\\rule{0ex}{1cm}}}\llap{\parbox[b]{6.cm}{$2$\\\rule{0ex}{1.35cm}}}\llap{\parbox[b]{6.5cm}{$1$\\\rule{0ex}{0.75cm}}}
    \includegraphics[width=.24\columnwidth]{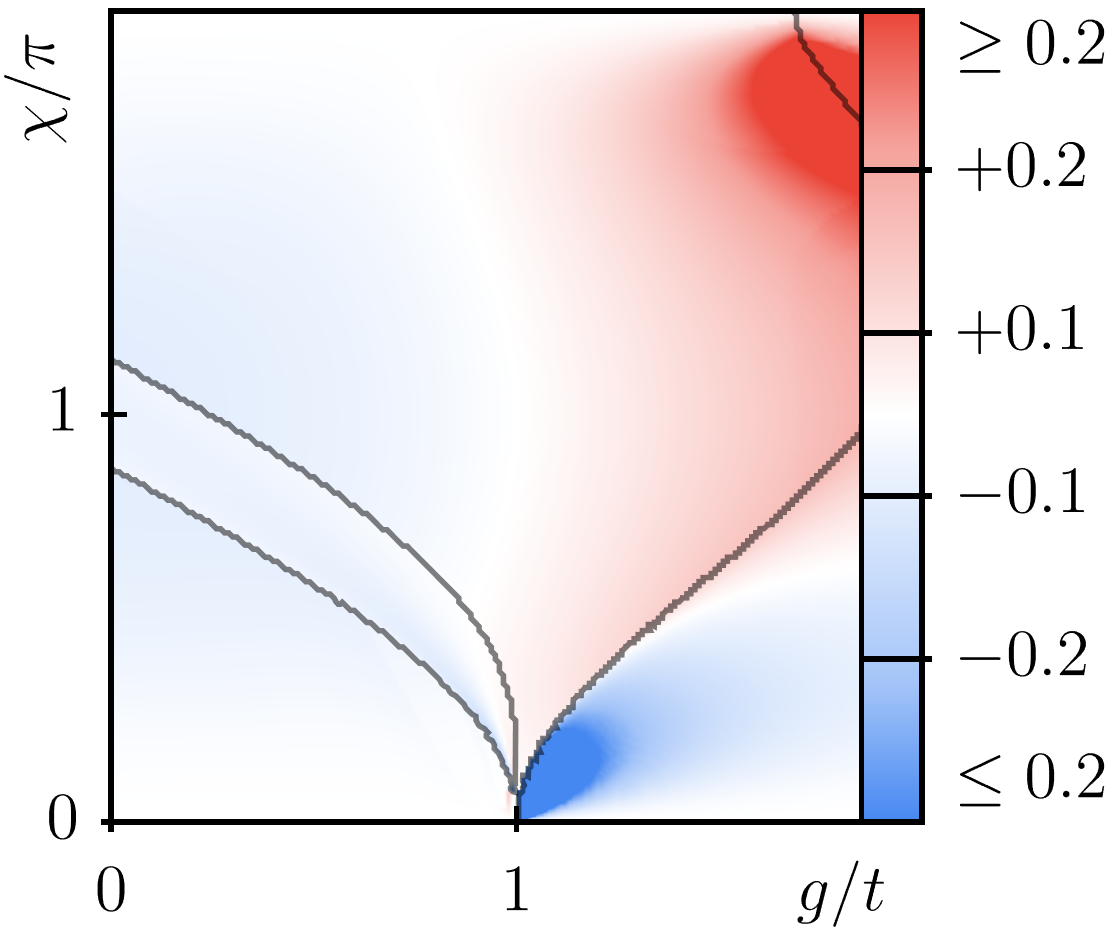}\llap{\parbox[b]{6.8cm}{(d)\\\rule{0ex}{2.95cm}}}\llap{\parbox[b]{4.8cm}{$c=1$\\\rule{0ex}{2cm}}}\llap{\parbox[b]{2.8cm}{$2$\\\rule{0ex}{1cm}}}\llap{\parbox[b]{6.cm}{$2$\\\rule{0ex}{1.35cm}}}\llap{\parbox[b]{6.5cm}{$1$\\\rule{0ex}{0.75cm}}}
    \caption{Change of the Drude weight by interaction according to first order perturbation theory in the thermodynamic limit for a system at incommensurate total density $n=20/32$ and $m/t=1$. (a) $\mathcal D_\parallel/V_\parallel$, (b) $\mathcal D_\perp/V_\perp$, (c) $\mathcal D_\parallel/V_\parallel+\mathcal D_\perp/V_\perp$ (d) $\mathcal D_U/U$. Solid lines mark the Lifshitz phase transitions between central charge $c=1$ and $c=2$. Fig.~\ref{fig:pt_mps} displays cuts of the density plots at $c=1$ close to $g/t=\chi/\pi=1$ with markers representing MPS results at $V_\perp/t = 0.1t$, $V_\parallel/t = 0.1t$ and $U/t = 0.125t$.}
    \label{fig:pert_results}
\end{figure}

We start by giving the first order correction of a generic density-density interaction of the form $\mathcal H_{\alpha,\beta,i} = V_{\alpha,\beta,i}\sum_j n_{\alpha,j}n_{\beta,j+i}$.
According to the Wick-theorem, we expand its two-point correlator following the usual contraction rules
\begin{align}
    \mathcal H_{\alpha,\beta,i} / V_{\alpha,\beta,i}
        =
    \sum_j\langle n_{\alpha,j}n_{\beta,j+i}\rangle
        =
    \sum_j
    \langle
        c^\dag_{\alpha,j}c^{\phantom\dag}_{\alpha,j}
        c^\dag_{\beta,j+i}c^{\phantom\dag}_{\beta,j+i}
    \rangle
        =
    \frac1L
    \sum_{k,l}
    \left(
    \langle
        n_{\alpha,k}
    \rangle
    \langle
        n_{\beta,l}
    \rangle
    +
    \langle
        c^\dag_{\alpha,k}c^{\phantom\dag}_{\beta,l}
    \rangle
    \langle
        c^{\phantom\dag}_{\alpha,l}c^\dag_{\beta,k}
    \rangle
    \cos((k-l)i)
    \right)
\end{align}
in which we assumed $n_{\alpha,j}n_{\beta,j+i} = 1/2 n_{\alpha,j}n_{\beta,j+i} + \hc$, i.e. any density-density correlator is a real function.
The expressions for the interactions of interest are now obtained by using the following single-particle expectation values
\begin{align}
    \label{eq:spev1}
    \langle c^\dag_{\uparrow,k}c^{\vphantom\dag}_{\uparrow,k}\rangle &= u(k)^2 \langle n_{+,k}\rangle + v(k)^2 \langle n_{-,k}\rangle\,,\\
    \label{eq:spev2}
    \langle c^\dag_{\downarrow,k}c^{\vphantom\dag}_{\downarrow,k}\rangle &= v(k)^2 \langle n_{+,k}\rangle + u(k)^2 \langle n_{-,k}\rangle\,,\\
    \label{eq:spev3}
    \langle c^\dag_{\uparrow,k}c^{\vphantom\dag}_{\downarrow,k}\rangle &= + u(k)v(k) \langle n_{+,k}\rangle - u(k)v(k) \langle n_{-,k}\rangle\,.
\end{align}
The interactions considered in the main text are simply given by $\mathcal H_U = \mathcal H_{\uparrow,\downarrow,0}$, $\mathcal H_\parallel = \sum_\alpha\mathcal H_{\alpha,\alpha,1}$ and $\mathcal H_\parallel = \mathcal H_{\uparrow,\downarrow,1} + \mathcal H_{\downarrow,\uparrow,1}$.
By projection onto the lower band, i.e. $n_{+,k}=0$, and substituting the expressions in Eqs.~\eqref{eq:spev1}-\eqref{eq:spev3} for the spinful densities, we find
\begin{align}
    \mathcal H_\parallel/2 &= \frac {V_\parallel}L\sum_{k,l}\left(u_k^2 v_l^2 - u_k^2(1-v_l^2)\cos(k-l)\right)\langle n_{-,k}\rangle\langle n_{-,l}\rangle\,,\\
    \mathcal H_\perp/2 &= \frac {V_\perp}L\sum_{k,l}\left(u_k^2 v_l^2 - u_kv_ku_lv_l\cos(k-l))\right)\langle n_{-,k}\rangle\langle n_{-,l}\rangle\,\\
    \mathcal H_U &= \frac{U}L\sum_{k,l}\left(u_k^2 v_l^2 - u_kv_ku_lv_l\right)\langle n_{-,k}\rangle\langle n_{-,l}\rangle\,.
\end{align}
We use the more convenient form of the coherence factors which relates their squares to the components of the underlying Hamiltonian's Bloch vector components by the following relations
\begin{align}
    u_k^2 &= \frac12(1+\tilde h_z)
    &
    v_k^2 &= \frac12(1-\tilde h_z)
    &
    2u_kv_k &= \tilde h_x
    \,.
\end{align}
to arrive at
\begin{align}
    \mathcal H_\parallel &= \frac {V_\parallel}{2L}\sum_{k,l}\left(1 + \tilde h_z(k)\tilde h_z(l) - \tilde h_z(k)\tilde h_z(l)\cos(k-l))\right)\langle n_{-,k}\rangle\langle n_{-,l}\rangle\,,\\
    \mathcal H_\perp &= \frac{V_\perp}{2L}\sum_{k,l}\left(1 - \tilde h_z(k)\tilde h_z(l) - \tilde h_x(k)\tilde h_x(l)\cos(k-l))\right)\langle n_{-,k}\rangle\langle n_{-,l}\rangle\,,\\
    \mathcal H_U &= \frac{U}{4L}\sum_{k,l}\left(1 - \tilde h_z(k)\tilde h_z(l) - \tilde h_x(k)\tilde h_x(l)\right)\langle n_{-,k}\rangle\langle n_{-,l}\rangle\,.
\end{align}
Coupling with a magnetic field penetrating the ring of atoms is done by introducing a twist $\re^{\ri\Phi}$ in the boundary conditions and, ultimately, results in substituting the arguments in the sums by $k\rightarrow k-\Phi/L$ as written in the main text.
The final expansion in $\Phi/L$ can be done easily by using the following set of equations
\begin{align}
    \tilde h_i(k-\Phi/L) &\approx \tilde h_i(k) - \frac\Phi L \tilde h_i'(k) + \frac{\Phi^2}{2L^2} \tilde h_i''(k)\,,\\
    \left(\sum_k\tilde h_i(k-\Phi/L)\right)^2 &\approx \left(\sum_k\tilde h_i(k)\right)^2 + \frac{\Phi^2}{L^2} \left(\left[\sum_k \tilde h'_i(k)\right] + \sum_k\tilde h_i''(k)\sum_l h_i(l)\right)\,,\\
    \sum_{k,l}\tilde h_i(k-\Phi/L)\tilde h_i(l-\Phi/L)\cos(k-l) &\approx \sum_{k,l}\cos(k-l)\left(\tilde h_i(k)\tilde h_i(l)+\frac{\Phi^2}{L^2}\left[\tilde h_i''(k)\tilde h_i(l) + \tilde h_i'(k)\tilde h_i'(l)\right]\right)\,,\\
    \sum_{k,l}\tilde h_x(k-\Phi/L)\tilde h_x(l-\Phi/L)\cos(k-l) &\approx \left(\sum_k\cos(k)\tilde h_x\right)^2 + \frac{\Phi^2}{L^2}\left(\left[\sum_k\sin(k)\tilde h_x'(k)\right]^2 + \sum_{k,l}\cos(k)\cos(l)\tilde h_x''(k)\tilde h_x(l)\right)\,,\\
    \sum_{k,l}\tilde h_z(k-\Phi/L)\tilde h_z(l-\Phi/L)\cos(k-l) &\approx \left(\sum_k\sin(k)\tilde h_z\right)^2 + \frac{\Phi^2}{L^2}\left(\left[\sum_k\cos(k)\tilde h_z'(k)\right]^2 + \sum_{k,l}\sin(k)\sin(l)\tilde h_z''(k)\tilde h_z(l)\right)\,,
\end{align}
in which we used the fact that odd orders in $\Phi$ vanish.
Therefore, all the above equations are exact up to $\mathcal O(\Phi^4)$.
Note that in the above, the sums in momentum space are considering all occupied momenta of the bottom band.
Finally, the notion of integrals of the Bloch vector components $H^{\alpha,\beta}_{i,n}$ yields the equations discussed in the main text.
To visualize the results, we plot the Drude weight shift of each interaction $\mathcal D_\parallel/V_\parallel$, $\mathcal D_\perp/V_\perp$ and $\mathcal D_U/U$ in Fig.~\ref{fig:pert_results}.

\section{Bosonization -- supplementary}
\label{app:bosonization_details}
Here we present the detailed derivation of the bosonized Hamiltonian in Eq.~\eqref{eq:ll} of the Creutz model.
We begin by projecting the spinors onto left- and right-moving fermions of the bottom band.
\begin{align} \label{eq:linearized_spinors}
c_{\uparrow,j} &= -e^{ik_{\rm F}j}v(k_{\rm F})R_j - e^{-ik_{\rm F}j}v(-k_{\rm F})L_j \,,
&
c_{\downarrow,j} &= +e^{ik_{\rm F}j}u(k_{\rm F})R_j + e^{-ik_{\rm F}j}u(-k_{\rm F})L_j \,.
\end{align}
Next, we insert the projecton into the spatial spin-densities in the continuum
\begin{align} \label{eq:effective_density_up}
    n_\uparrow(x) &= v^2(k_{\rm F}) n_R(x) + v^2(-k_{\rm F}) n_L(x) + v(-k_{\rm F})v(k_{\rm F})\Big[e^{-2ik_{\rm F}x}R^\dag(x) L(x) +\mbox{h.c}\Big]\,,\\
    \label{eq:effective_density_down}
    n_\downarrow(x) & = u^2(k_{\rm F}) n_R(x) + u^2(-k_{\rm F}) n_L(x) + u(k_{\rm F})u(-k_{\rm F})\Big[e^{-2ik_{\rm F}x}R^\dag(x) L(x) +\mbox{h.c.}\Big]\,,
\end{align}
with $x=ja$ and $a$ a dimensional lattice spacing.
Next, we plug the densities readily into the expressions of the interactions of interest.
One thus finds the effective low-energy expressions according to
\begin{align}
    \begin{split}
    \label{eq:hVlin} \mathcal H_\parallel+\mathcal H_\perp=\sum_x\Big[&\gamma_1\Big(n_R(x)n_R(x+a)+n_L(x)n_L(x+a)\Big)+\gamma_2\Big(n_R(x)n_L(x+a)+n_L(x)n_R(x+a)\Big)\\
    &\qquad+\gamma_3\Big(e^{2ik_{\rm F}a}R^\dagger(x)L(x)L^\dagger(x+a)R(x+a)+\mbox{h.c.}\Big)\Big]\,.
    \end{split}
\end{align}
\begin{align}
\gamma_1&=(u^4+v^4)V_\parallel+ 2u^2v^2V_\perp\,, &\gamma_2&=2u^2v^2V_{\parallel}+(u^4+v^4)V_\perp\,, & \gamma_3&=2u^2v^2(V_\parallel+V_\perp)\,,
\end{align}
in which we used the shortand notation $u/v=u(k_{\rm F})/v(k_{\rm F})$.
Finally, we find the bosonized Hamiltonian by applying the standard bosonization identities
\begin{align}
    R(x) &\sim
    \frac1{\sqrt{2\pi a}} \re^{-\ri (\phi(x)-\theta(x))}\,,
    &
    L(x) &\sim
    \frac1{\sqrt{2\pi a}} \re^{\ri (\phi(x)+\theta(x))}
    \,,
    \label{eq:spinless_bosonization_identity}
\end{align}
leading to the bosonization of the point-split operator
\begin{align}\label{eq:point_split}
R^\dag(x)L(x)L^\dag(x+a)R(x+a) + \hc = -\frac{2\partial_x\phi^2}{(2\pi)^2} \,,
\end{align}
in which we neglect some infinite, but constant, terms.
We notice this expression only involves the density field $\phi$.
Thus, it does not affect the current operator and has no effect on the Drude weight, even though it affects $g_4$ and $g_2$.
Finally, we remind that the mover-densities relate to the bosonic fields according to
\begin{align}
    \label{eq:bi}
    n_R(x)&= [\partial_x\theta(x)-\partial_x\phi(x)]/2\pi\,,
    &
    n_L(x)&=-[\partial_x\theta(x)+\partial_x\phi(x)]/2\pi\,.
\end{align}
Following this recipe and plugging Eqs.~\eqref{eq:bi} into Eq.~\eqref{eq:hVlin} yields the coupling constants given in Tab.~\ref{tab:copuling_constants} which concludes this supplementary section.
\begin{table}[ht]
    \centering
    \begin{tabular}{c c c}
        \toprule
        $V_i$ & $g_{4,i}$ & $g_{2,i}$\\
        \midrule
        $V_\parallel$ & $2(u^4+v^4)-4u^2v^2\cos(2k_{\rm F})$ & $g_{4,\perp}$\\
        $V_\perp$ & $4u^2v^2(1-\cos(2k_{\rm F}))$ & $g_{4,\parallel}$\\
        \bottomrule
    \end{tabular}
    \caption{Renormalization of the Luttinger parameters in Eq.~\eqref{eq:uk}. The first column lists the different nearest-neighbor interactions and corresponding $g_2$ and $g_4$ factors.} \label{tab:copuling_constants}
\end{table}

\section{Crossover to two-dimensional systems}\label{app:multi}
\begin{figure}[tt]
    \centering
    \includegraphics[width=.249\textwidth]{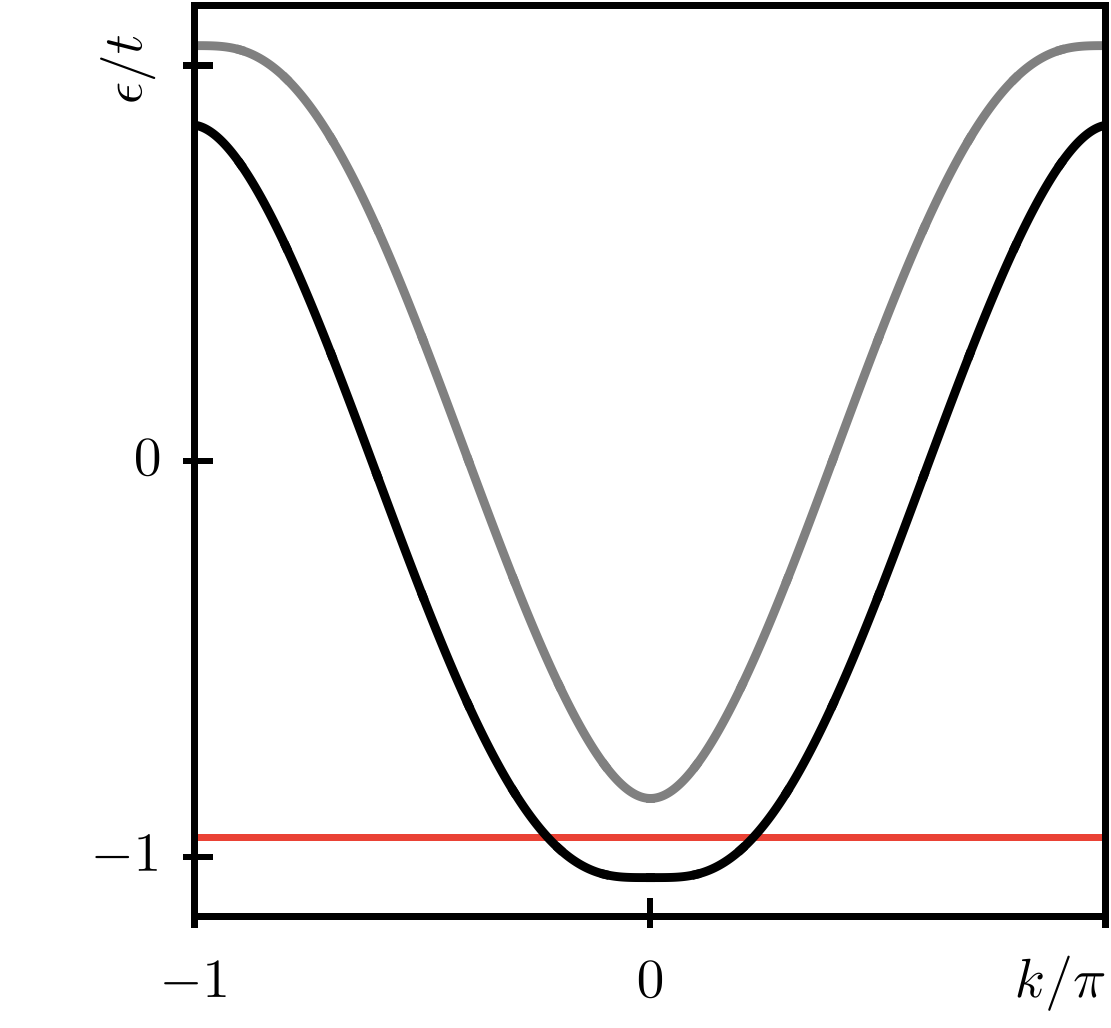}\llap{\parbox[b]{3.7cm}{(a)\\\rule{0ex}{3.6cm}}}
    \hfil
    \includegraphics[width=.249\columnwidth]{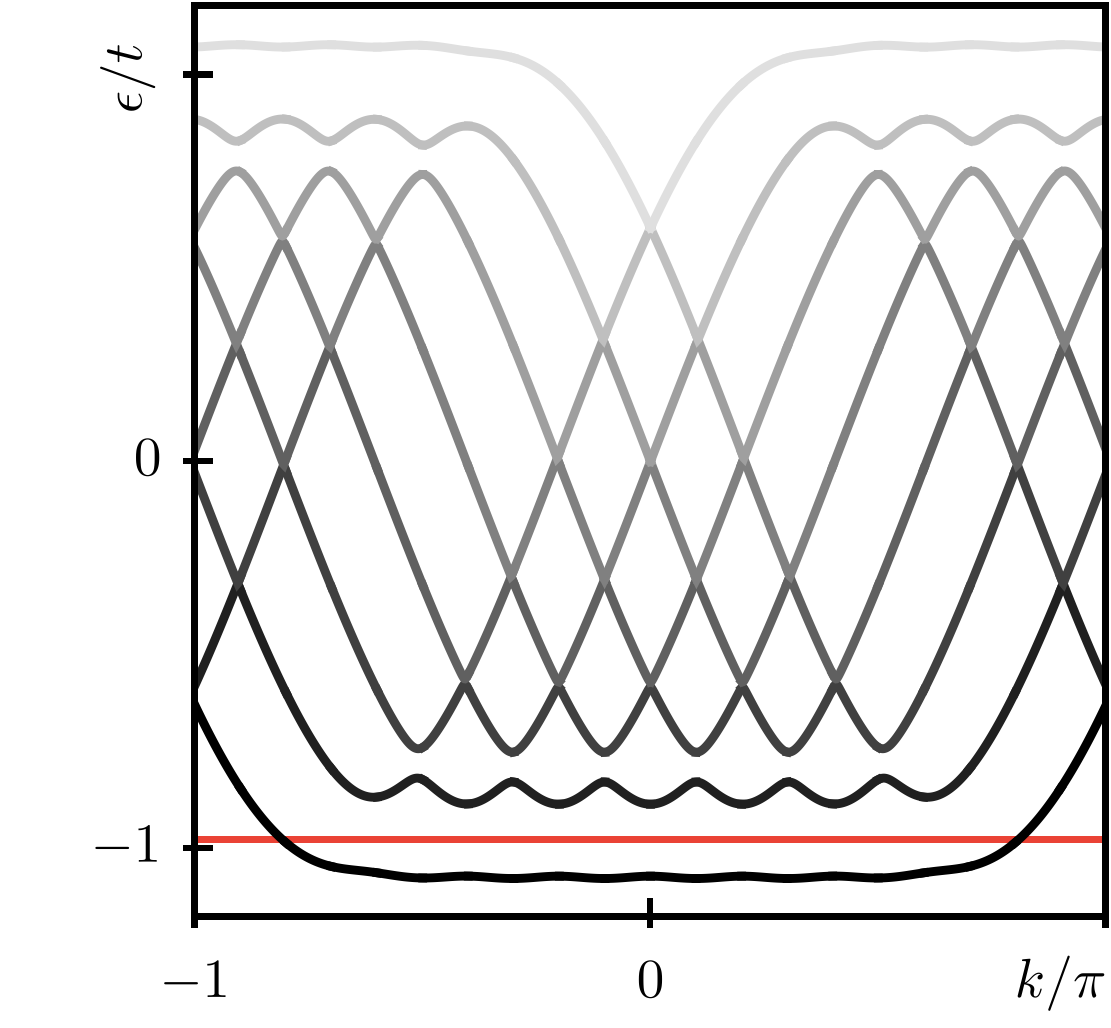}\llap{\parbox[b]{3.7cm}{(b)\\\rule{0ex}{3.6cm}}}
    \hfil
    \includegraphics[width=.249\columnwidth]{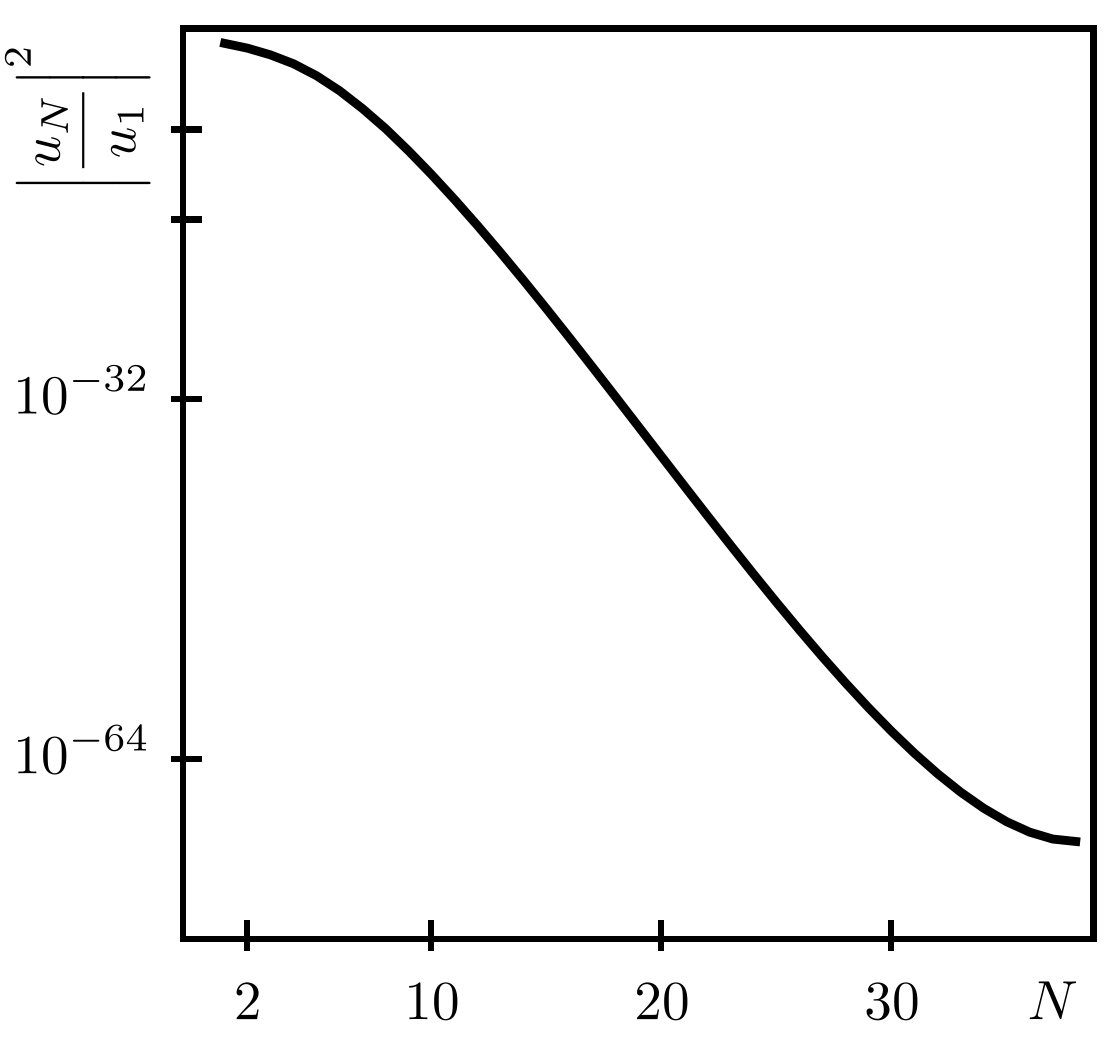}\llap{\parbox[b]{3.7cm}{(c)\\\rule{0ex}{3.6cm}}}
    \caption{Coupled wire approach with (a) $N=2$ and (b) $N=7$. We place the Fermi energy in the center of the gap, marked as a red line. In (c), we plot the amplitude $\left|u_N/u_1(k_{\rm F})\right|^2$ for the case $m=.1t$, $g=0$ and $\chi=0.025\pi$ versus number of chains $N$, evidently following an exponential decay. Following Tab.~\ref{eq:uk} this implies a strong suppression of $g_2$ processes compared to $g_4$ in the $c=1$ gap.}
    \label{fig:coupled_wires}
\end{figure}

We generalize the model defined in Ch.~\ref{section:interacting_creutz} to the case of $N$ legs of the quasi one-dimensional wire.
In momentum space, we choose the transverse flux $\chi$ such that the generalized Hamiltonian takes the simple form
\begin{align}
    H_0^N(k) = \sum_{j_y} \left(m-g\cos k\right)(c^\dag_{j_y\vphantom{+1},k} c^{\vphantom\dag}_{j_y+1,k}+{\rm h.c.}) - t\cos\left(k-\chi (2j_y-(N+1))\right)n_{j_y,k}\,.
\end{align}
This Hamiltonian may be diagonalized by a rotation onto the bands $U_N(k)$ such that $U_N(k) H_0^N(k) U_N^\dag(k) = {\mathcal E}_N(k)$, with a diagonal matrix ${\mathcal E}_N(k) = {\rm diag}(\varepsilon_1(k),\varepsilon_2(k),\ldots,\varepsilon_N(k))$ containing the ordered bands $\varepsilon_i(k) \leq \varepsilon_{i+1}(k)$ (see Fig.~\ref{fig:coupled_wires}).
Although more general scenarios may be considered, we focus here on the coupled wire formalism which is obtained by setting $g=0$.
In particular, this adresses the one-dimensional limit of the Harper-Hofstadter model~\cite{harper_single_1955,hofstadter_energy_1976}, understood as a coupled-wire system~\cite{kane_fractional_2002} and studied recently also in the context of superconductivity~\cite{yang_from_topology_to_superconductivity_2019}.
We require a central charge $c=1$ such that one has a single pair of left and right moving species defined on the lowest band $\varepsilon_1(\pm k_{\rm F})$.
This justifies the approximation similar to Eq.~\ref{eq:linearized_spinors},
\begin{align} \label{eq:linearized_spinors_general}
    \begin{pmatrix}
        c_{1,x} \\
        c_{N,x}
    \end{pmatrix}
    \sim
    \begin{pmatrix}
        u_1(k_{\rm F}) & u_1(-k_{\rm F}) \\
        u_N(k_{\rm F}) & u_N(-k_{\rm F})
    \end{pmatrix}
    \begin{pmatrix}
        R(x)\re^{+\ri k_{\rm F} x} \\
        L(x)\re^{-\ri k_{\rm F} x}
    \end{pmatrix}
\end{align}
with some prefactors $u_1$ and $u_N$ which depend on the precise form of $U_N$.
A more general form of the imposed interactions in the N-legged setup is
\begin{align}
    \mathcal H_\parallel = \sum_{j_x,j_y}V_\parallel(j_y)n_{j_x,j_y}n_{j_x+1,j_y}
\end{align}
where the local leg density is defined as $n_{j_x,j_y} = c^\dag_{j_x,j_y}c^{\phantom\dag}_{j_x,j_y}$.
Due to the nature of the applied projection, other interactions similar to $\mathcal H_\perp$ coupling adjacent wires along the $y$-direction will not contribute to the effective model.
A hypothetical interaction similar to the two-wire case which contributes to the projected Hamiltonian is
\begin{align}
    \mathcal H_\perp = V_\perp(N) \sum_{j_x} n_{j_x,1}n_{j_x+1,N}\,.
\end{align}
The remaining calculation to arrive at the Luttinger Liquid Hamiltonian is similar to the one presented in main text, and, for $V_\parallel(1)=V_\parallel(N)$ identical $g$-factors are recovered (substituting $u\rightarrow u_1$, $v\rightarrow u_N$, $V_\parallel\rightarrow V_\parallel(1)$ and $V_\perp\rightarrow V_\perp(N)$).
As a consequence, the result here is analog to the one presented in the main text
\begin{equation}
g_{4,\parallel}-g_{2,\parallel}=2(u_1^2-u_N^2)^2>0\,.
\end{equation}

In Fig.~\ref{fig:coupled_wires} we explicitly target the 2D limit $N\rightarrow\infty$, showing that $u_1$ is the leading contribution -- thus strongly suppressing $g_{2,\parallel}$ processes in the effective model.
It has to be stressed that Eq.~\ref{eq:linearized_spinors_general} is not a unitary transformation and projects onto the relevant low-energy sector of the model.
In the case of $N>2$, the bulk is fully projected out in the effective model and only effects on the edges of the system can be recovered.
This makes it impossible to deduce bulk properties like the Hall conductivity $\sigma_H$ that is instead universally quantized at a given filling.
Instead, what we claim here is a strong modification of the edge Drude weight due to interactions for which a universal scaling law dependent on the interactions has been reported using chiral Luttinger liquid approaches \cite{antinucci_universal_2018}.
Building up on this statement, we provide here a generic example that enhancement effects of the Drude weight become universally applicable for systems hosting polarized conduction bands dressed with repulsive density-density interactions.

\section{Comparison between leading order perturbation theory and bosonization}\label{app:pt_vs_bosonization}
\begin{figure}[ht] \label{fig:rotation_derivatives}
    \centering
    \includegraphics[width=.24\columnwidth]{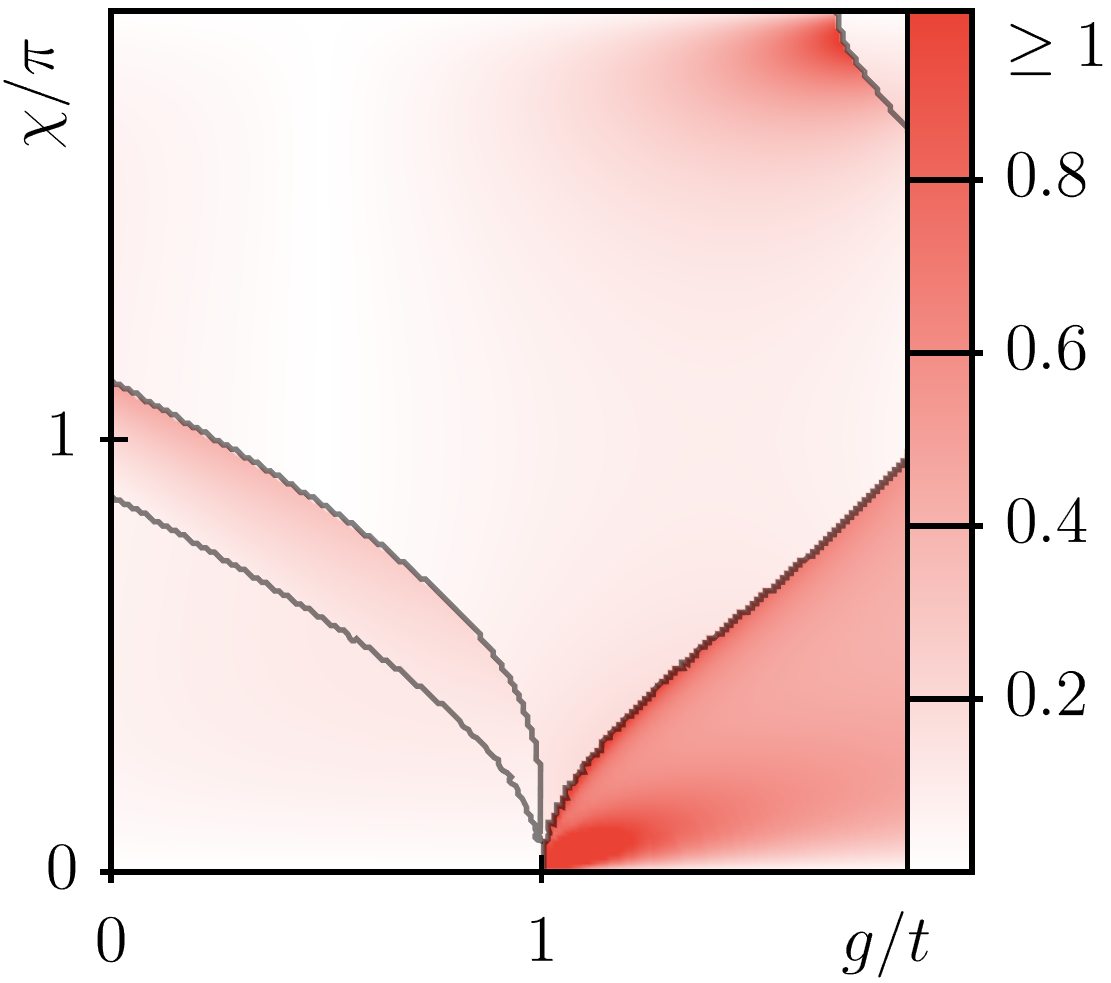}\llap{\parbox[b]{6.8cm}{(a)\\\rule{0ex}{2.95cm}}}\llap{\parbox[b]{4.8cm}{$c=1$\\\rule{0ex}{2cm}}}\llap{\parbox[b]{2.8cm}{$2$\\\rule{0ex}{1cm}}}\llap{\parbox[b]{6.cm}{$2$\\\rule{0ex}{1.48cm}}}\llap{\parbox[b]{6.5cm}{$1$\\\rule{0ex}{0.75cm}}}
    \includegraphics[width=.24\columnwidth]{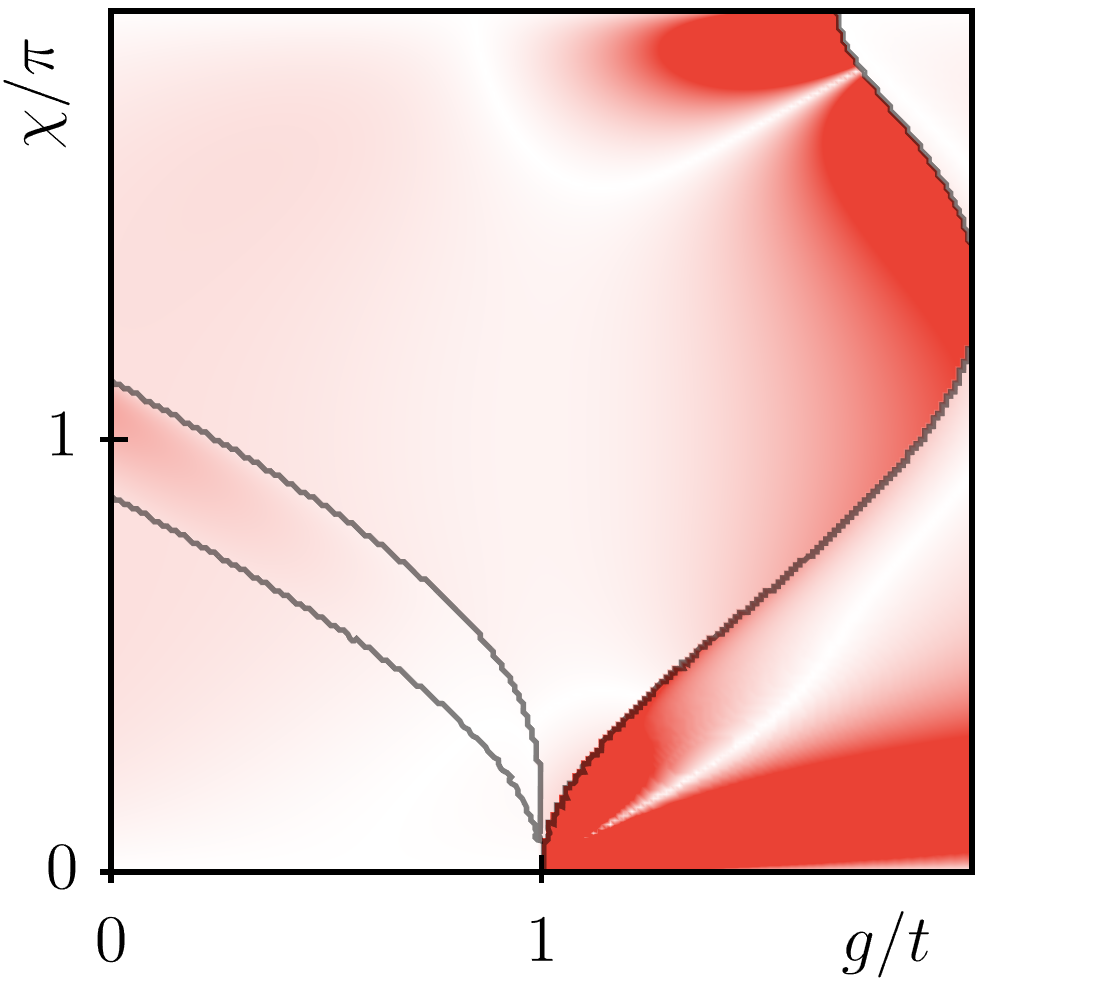}\llap{\parbox[b]{6.8cm}{(b)\\\rule{0ex}{2.95cm}}}\llap{\parbox[b]{4.8cm}{$c=1$\\\rule{0ex}{2cm}}}\llap{\parbox[b]{2.8cm}{$2$\\\rule{0ex}{1cm}}}\llap{\parbox[b]{6.cm}{$2$\\\rule{0ex}{1.48cm}}}\llap{\parbox[b]{6.5cm}{$1$\\\rule{0ex}{0.75cm}}}
    \includegraphics[width=.24\columnwidth]{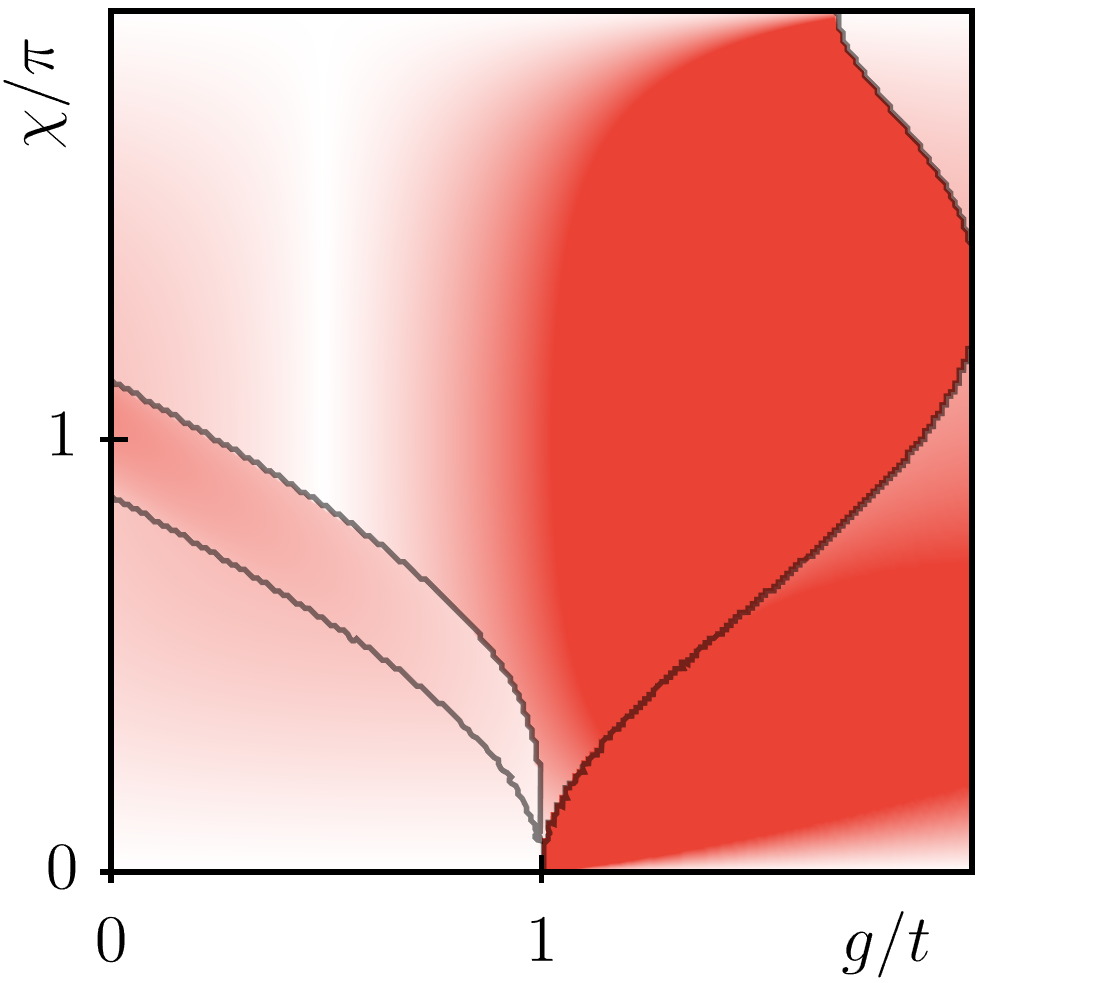}\llap{\parbox[b]{6.8cm}{(c)\\\rule{0ex}{2.95cm}}}\llap{\parbox[b]{4.8cm}{$c=1$\\\rule{0ex}{2cm}}}\llap{\parbox[b]{2.8cm}{$2$\\\rule{0ex}{1cm}}}\llap{\parbox[b]{6.cm}{$2$\\\rule{0ex}{1.48cm}}}\llap{\parbox[b]{6.5cm}{$1$\\\rule{0ex}{0.75cm}}}
    \includegraphics[width=.24\columnwidth]{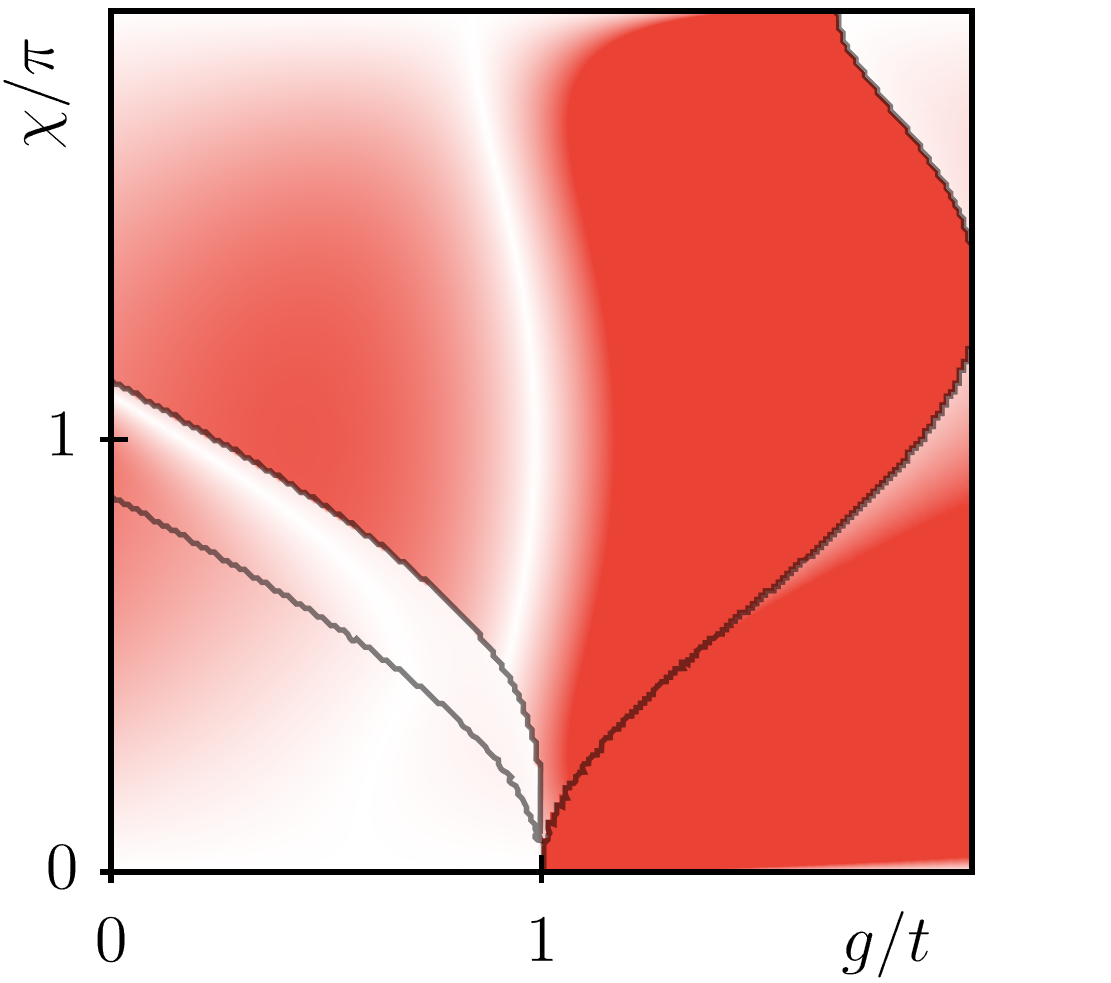}\llap{\parbox[b]{6.8cm}{(d)\\\rule{0ex}{2.95cm}}}\llap{\parbox[b]{4.8cm}{$c=1$\\\rule{0ex}{2cm}}}\llap{\parbox[b]{2.8cm}{$2$\\\rule{0ex}{1cm}}}\llap{\parbox[b]{6.cm}{$2$\\\rule{0ex}{1.48cm}}}\llap{\parbox[b]{6.5cm}{$1$\\\rule{0ex}{0.75cm}}}
    \caption{(a)-(b) First two derivatives of $\left|u^{(n)}/u(k_{\rm F})\right|$, (c)-(d) first two derivatives of $\left|v^{(n)}/v(k_{\rm F})\right|$ versus $g$ and flux $\chi$ at density $n=2/3$ and $m=1$. Discrepancies between bosonization and perturbation arise due to a strong violation of Eq.~\eqref{eq:local_mover_reqs}.}
\end{figure}

Major deviations between perturbation theory and bosonization arise from Eq.~\ref{eq:linearized_spinors} which requires a flat rotation matrix $U$.
In particular, according to first order perturbation theory, the canonical operators can be expressed using the bottom band modes only, i.e.
\begin{align}
    c_{\uparrow,j} &= \re^{\ri k_{\rm F}x}v(k_{\rm F})\sum_{q}\re^{\ri qx}\left(1+\sum_{n=1}^\infty \frac{q^n}{n!}\frac{v^{(n)}}v(k_{\rm F})\right)d_{-,q+k_{\rm F}} + \re^{-\ri k_{\rm F}x}v(-k_{\rm F})\sum_{q}\re^{\ri qx}\left(1+\sum_{n=1}^\infty \frac{q^n}{n!}\frac{v^{(n)}}v(-k_{\rm F})\right)d_{-,q-k_{\rm F}}
\end{align}
and a similar expression ($v\leftrightarrow u$) holds for the down-species.
To arrive at Eq.~\ref{eq:linearized_spinors} in the main text, we further assume
\begin{align}\label{eq:local_mover_reqs}
    \sum_{n=1}^\infty\frac{q^n}{n!}\frac{f^{(n)}}f(\pm k_{\rm F}) \approx 0\,,\ f\in\{u,v\}\,.
\end{align}
The momentum transfer in forward-scattering processes $g_4$ is small compared to the Fermi momentum and the above equation is automatically satisfied.
However, back-scattering processes $g_2$ always transfer momentum comparable to $k_{\rm F}$ and the above equation requires the rotation matrix being almost constant in $q$, i.e. $U^{(n)}/U(q)\approx 0$.
We show a few example contours of the first three derivatives of $\left|f^{(n)}/f(k_{\rm F})\right|$ along $g$ and $\chi$ in Fig.~\ref{fig:rotation_derivatives} which indeed explains major deviations at regions in which $u^{(n)}/u$ and $v^{(n)}/v$ are large.
On the contrary, at the single Dirac cone point $g=\chi/\pi=1$ the rotation matrix has entries $u/v(k)=\cos/\sin (k/4)$, as a consequence the $n$'th derivatives are exponentially suppressed in $n$ and Eq.~\ref{eq:local_mover_reqs} becomes a reasonable approximation.
Clearly, in regions outside of Eq.~\ref{eq:local_mover_reqs}, $g_2$ processes are not correctly considered in the effective model and thus strong deviations in $\mathcal D_\perp$ are expected.
Remarkably, the qualitative results of the effective model stay valid at the full region of $c=1$, i.e. we find a positive shift of the Drude weight induced by $\mathcal H_\parallel$, and, a negative effect of comparable amplitude due to $\mathcal H_\perp$.
In conclusion, the above equation fully disregards band- and coherence factor curvature, yielding a consistent result with leading order perturbation theory up to zeroth order in the coherence factor derivatives only (${H_{z,1}^{0,0}} = 2 {{\tilde h}_z(k_{\rm F})}$ terms in the Drude weights are recovered).
For a future investigation, it might be interesting to relax Eq.~\ref{eq:local_mover_reqs}, in particular including higher order terms in $q$.
This not only accounts for a sum of right and left movers on {\em different} lattice sites in Eq.~\ref{eq:linearized_spinors}, it naturally reintroduces non-trivial curvature of the coherence factors $u(k)$ and $v(k)$, ultimately giving subleading corrections to $\gamma_i$ in Eq.~\ref{eq:hVlin}.

\section{Details on the MPS simulations}\label{app:mps_quality}
The matrix product states (MPS) results presented in the main text were performed by our own implementation of a U(1) symmetric code preserving the total number of particles, based on the anthology of tensor networks build on a symmetry-preserving library in collaboration with the group of S. Montangero at the University of Ulm~\cite{silvi_mps}.

\begin{figure}[ht]\label{fig:mps_quality}
    \centering
    \includegraphics[height=3.6cm]{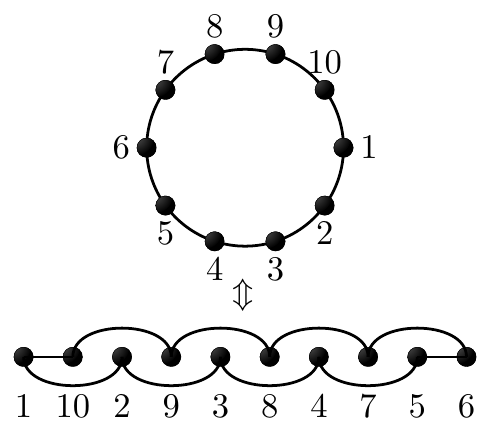}\llap{\parbox[b]{7cm}{(a)\\\rule{0ex}{1cm}}}
    \includegraphics{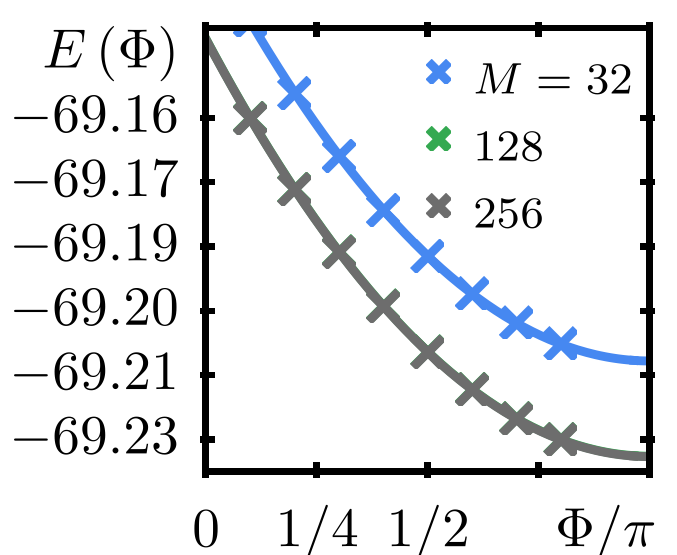}\llap{\parbox[b]{5cm}{(b)\\\rule{0ex}{.6cm}}}
    \includegraphics{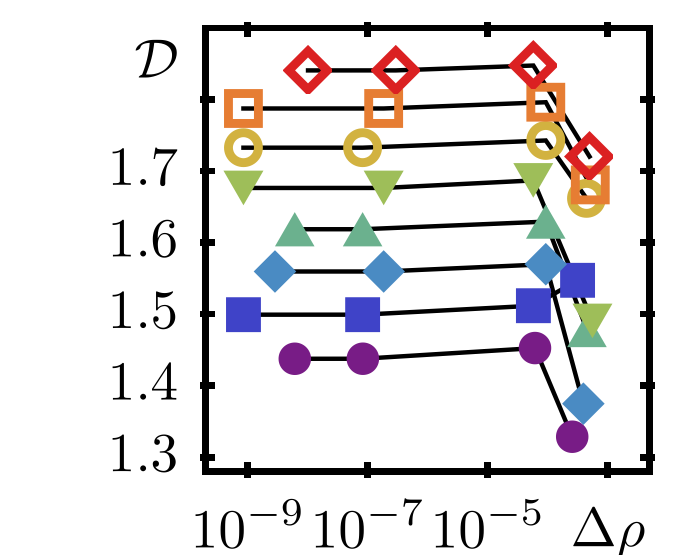}\llap{\parbox[b]{5cm}{(c)\\\rule{0ex}{.6cm}}}
    \includegraphics[height=3.6cm]{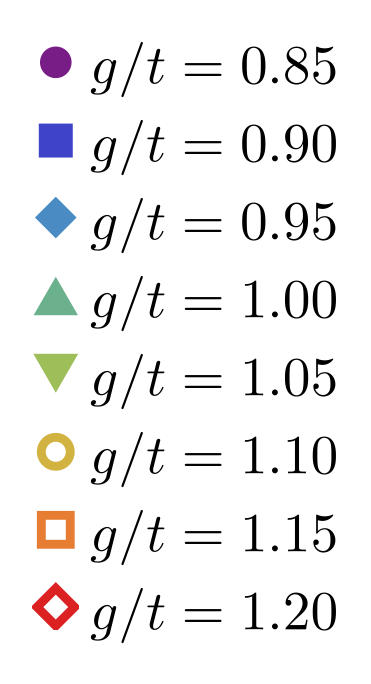}
    \caption{(a) Deformation of a generic ring system to simulate periodic boundary conditions with open boundary MPS algorithms. (b) Generic representative of numerical simulations with different bond dimension $M$ of the groundstate energy dependence $E(\Phi)$ for a system at fixed density $n=N/L=20/32$ at parameters $m/t=\chi/\pi=1$, $g/t=1.2$, $U/t=4$ and $V_\parallel=V_\perp=0$. Crosses represent MPS results and continuous lines represent a quadratic fit. At the presented decimal precision, data and fits for $M=128$ and $M=256$ reside exactly on top of each other. (c) Dependence of the Drude weight on the truncated probability $\Delta\rho$ of the MPS. In case of $\Delta\rho<10^{-7}$, the data is sufficiently converged up to the necessary precision presented in the main text and the convergence error may be disregarded.}
\end{figure}

Periodic boundary conditions (PBC) in general are hard to tackle using tensor network schemes due to the absence of a canocial form, which is a necessity to simplify the computational complexity in variational optimizations~\cite{schollwock_density-matrix_2011}.
A naive solution used heavily in exact diagonalization is the introduction of long-range terms mimicking periodic boundary conditions by coupling the edges of the system.
However, such a strategy is deemed to fail for MPS Ansätze because, by construction, long-range correlations are not captured sufficiently.
An alternative scheme is obtained by deforming the ring into a 1D open boundary system with short-range next-to-nearest neighbor couplings, depicted in Fig.~\ref{fig:mps_quality}, panel (a).
This deformation amends the strong asymmetry between short-range hoppings and long-range boundary terms by reshuffling the lattice sites.
The two depicted ring geometries are numerically equivalent and yield an efficient simulation of periodic boundary conditions with open boundary MPS algorithms at the cost of slightly increasing the matrix product operator (MPO) dimension.

The extraction of the particle mobility $\mathcal D$ is straightforward -- we simulated the ground state energy dependence $E\left(\Phi\right)$ on the magnetic flux $\Phi$ penetrating the ring.
Numerically, coupling to a magnetic flux is readily done by applying the local transformations $t\rightarrow t\re^{\ri\Phi/L}$ and $g\rightarrow g\re^{\ri\Phi/L}$ in the real-space kinetic Hamiltonian $\mathcal H_{\rm kin}$.
The susceptibility function can then be extracted by two equivalent procedures: i) by approximating the second derivative according to $\partial_\Phi^2E(\Phi)=\left(E(\Phi+\epsilon)+E(\Phi-\epsilon)-2E(\Phi)\right)/\epsilon^2+\mathcal O(\epsilon^4)$, sending $\epsilon\rightarrow0$ and then $\Phi\rightarrow\Phi_m$, or by ii) fitting the energy dependence. Hereby, $\Phi_m$ is the energy extremum which depends on the parity of the underlying system. In most cases, $\Phi_m=0/\pi$ for an odd/even number of fermions respectively~\cite{filippone_controlled_2018}. According to leading order perturbation theory, it is possible to extract the mobility $\mathcal D$ by a quadratic fit function $f$ to approximate the energy dependence $E(\Phi)$ according to $f(\Phi) = a + b (\Phi - \Phi_m)^2$ with $\Phi_m$ being the energy extremum, and the Drude weight being $\mathcal D = 2\pi L b$, which is correct up to $\mathcal O((\Phi-\Phi_m)^4)$ in the flux $\Phi$.
Due to an astonishing agreement between prediction and numerics, even for strong interaction amplitudes, we present a generic representative of the fitting procedure in Fig.~\ref{fig:mps_quality} panel~(b).
A basic estimation of the numerical accuracy is governed by panel (c), in which we display the convergence of the presented observables versus $\Delta\rho$ -- the figure of merit in MPS simulations representing the truncated probability of the reduced density-matrix for the canonical bipartition at the center of the chain.

\section*{References}

\bibliographystyle{apsrev4-1}
\bibliography{biblio}

\end{document}